%% file: main.tex
\input{packages}
\begin{document}

\newcommand{\longbar}[1]{\mkern5mu\overline{\mkern-5mu#1\mkern-5mu}\mkern5mu}

\newcommand{\fabio}[1]{{\color{blue}#1}}


\title{
Neuromorphic computing with optomechanical oscillators}

\newcommand{\Bordeaux}{Université de Bordeaux, CNRS, LOMA, UMR5798, F-33400 Talence, France}

\author{Andrea Gaspari\orcidlink{0009-0009-5874-0638}}%
    \email{andrea.gaspari@u-bordeaux.fr}
\affiliation{\Bordeaux}

\author{Rémi Avriller \orcidlink{0000-0003-1582-9500}}
\affiliation{\Bordeaux}
\affiliation{Université de Strasbourg, CNRS, Institut de Physique et Chimie des Matériaux de Strasbourg, UMR7504, F-67000 Strasbourg, France}

\author{Florian Marquardt \orcidlink{0000-0003-4566-1753}}
\affiliation{Institute for Theoretical Physics, Universität Erlagen-Nürnberg, Staudtstr. 7, 91058 Erlangen, Germany}

\author{Fabio Pistolesi \orcidlink{0000-0002-5897-0347}}
\affiliation{\Bordeaux}

\date{\today}

\begin{abstract}
The increasing resource demands of artificial neural networks have prompted the exploration of novel platforms better suited for machine learning. 
In this context, phase oscillators represent a promising candidate due to their intrinsic nonlinearity and their ability to exhibit collective synchronization when coupled together. 
In the present work, we investigate one such implementation: a network of optomechanical oscillators pumped in the blue--detuned regime to achieve self--sustained oscillations.
We propose a theoretical framework to describe their dynamics and demonstrate how such systems can be employed for neuromorphic computing. 
We discuss how they can be trained and analyze a platform, based on drum resonators, that could enable their physical implementation. 
Ultimately, the theoretical results obtained from modelling an XOR gate using 5 nodes in an all--to--all configuration are discussed.
\end{abstract}

\maketitle

\input{1_Introduction}
\input{2_Model}
\input{3_Neural_network_implementation_on_physical_systems}
\input{4_XOR_Implementation}
\input{5_Conclusion}
\newpage
\input{6_Acknoledgements}
\input{7_Data}

\clearpage
\input{appendix}

\bibliography{Bibliography}
\end{document}

%% file: packages.tex
\documentclass[%
aps,
reprint,
superscriptaddress,
nofootinbib,
amsmath,amssymb,
prx,
]{revtex4-2}

\usepackage{orcidlink}
\usepackage{bm}
\usepackage{graphicx}
\usepackage{array}

\usepackage{xcolor}
\usepackage[table]{xcolor}

%% file: 1_Introduction.tex
\section{\label{sec_introduction}Introduction}

Artificial neural networks \cite{LeCun2015} (ANNs) have seen remarkable progress in recent years, driven by advances in both theoretical foundations and practical implementations. These developments have enabled models of unprecedented scale and accuracy, capable of tackling highly complex tasks. At the same time, this rapid growth has introduced significant engineering challenges arising from the increasing number of nodes and trainable parameters. For state--of--the--art ANNs, solving the network dynamics and implementing training protocols on conventional von Neumann computers \cite{vonNeumann1945} have become extraordinarily energy--intensive, exposing fundamental limitations. Computers, as deterministic and error-free platforms, are fundamentally inefficient at executing inherently stochastic machine learning algorithms, leading to a waste of computing power. 
Additionally, the physical separation between processing and memory units necessitates constant data transfer, which consumes substantial resources, especially in large--scale networks. This architecture also introduces a bottleneck effect: when computational demands exceed the data transfer capacity, the entire process slows down.
In this context, attention has increasingly focused on finding new solutions able to reduce energy consumption, mitigate the von Neumann bottleneck and enhance parallelism. The field of neuromorphic computing \cite{Christensen2022, Markovic2020} seeks to address these challenges by proposing novel platforms better suited to machine learning.  
Notably, one of the main strategies adopts a genuinely physical approach, aiming to replace artificial neurons with dynamical systems and harness their intrinsic nonlinear behavior for computation. Many implementations have been already explored, including networks based on memristors \cite{Prezioso2015}, spin oscillators \cite{Torrejon2017, Romera2018} and Josephson junctions \cite{Schneider2018}, as well as optical proposals involving light scattering in complex media \cite{Gigan1, Gigan2}, multiple light scattering \cite{Yildirim2024}, multiple--scattering cavities \cite{Xia2024}, linear wave scattering \cite{Wanjura2024}, diffractive optical elements \cite{Bueno:18} and coherent nanophotonic circuits \cite{Shen2017}. 
However, it remains unclear which physical platforms are most suitable in terms of efficiency, scalability and expressivity. This motivates the exploration of further system classes--- ideally those with established communities and robust theoretical foundations.

Following the remarkable results obtained using phase oscillators \cite{Wang2024}, our attention was directed toward optomechanical implementations, broadly defined across the microwave to the optical regime. These systems exhibit a broad range of properties \cite{Aspelmeyer} that make them particularly attractive for neuromorphic computing. Specifically, they feature an intrinsic nonlinearity arising from the interaction with the cavity field (essential for nonlinear computing), micro-- and nanoscale physical dimensions, high quality factors and high oscillation frequencies, as well as the possibility of being cooled down to millikelvin temperatures to suppress thermal noise. Moreover, they offer precise tunability of the fabrication parameters, making them extremely versatile platforms for experimental implementations. In parallel, when the mechanical damping is balanced by a pumping source, these systems enter the self-sustained regime and can be effectively treated as phase oscillators. Under such condition, they undergo a Hopf bifurcation and start oscillating on stable limit orbits. Networks made of phase oscillators are particularly appealing for computing as they can display collective phenomena like synchronization \cite{Pikovsky_Rosenblum_Kurths_2001}— an emergent behavior, well documented in neuroscience, which we aim to exploit.

The structure of the paper is as follows. Starting from the theoretical model, in Sec.\ref{sec:_Optomechanical_nodes} and \ref{sec_Optomechanical network} the individual and collective dynamics of optomechanical oscillators is analyzed, developing a framework tailored for neuromorphic computing. The approach relies on using oscillator phases as node variables and leverages coupling, external driving and synchronization to reproduce the main features of artificial neural networks. In Sec.\ref{sec_Physical_ML}, the implementation of inference and training on physical platforms is presented, with a particular focus on the role of dynamical multistability discussed in Sec.\ref{Multistability}. To complete the analysis, in Sec.\ref{Realization of physical interconnections.} we present a model to physically realize the interconnections using drum resonators \cite{Teufel2011,Lenhert} while in Sec.\ref{sec:_Experimental feasibility} we delve into the case of SiN drum resonators \cite{Xin}, providing parameter ranges for the experimental realization. Ultimately, the entire scheme is validated through the theoretical implementation of an XOR gate (Sec.\ref{XOR_implementation}), an ideal benchmark in machine learning due to its simplicity. The results, collected in Sec.\ref{Results}, confirm the network's expressivity, \textit{i.e.} its ability to successfully solve the task.   


%% file: 2_Model.tex
\section{\label{sec_model}Model}
Starting from the physical description of a single optomechanical oscillator, this section gradually builds the full network model, placing particular emphasis on the analogies with ANNs.

\subsection{\label{sec:_Optomechanical_nodes}Optomechanical nodes} 

\begin{figure} [t]
    \centering
    \includegraphics[width=0.98\linewidth]{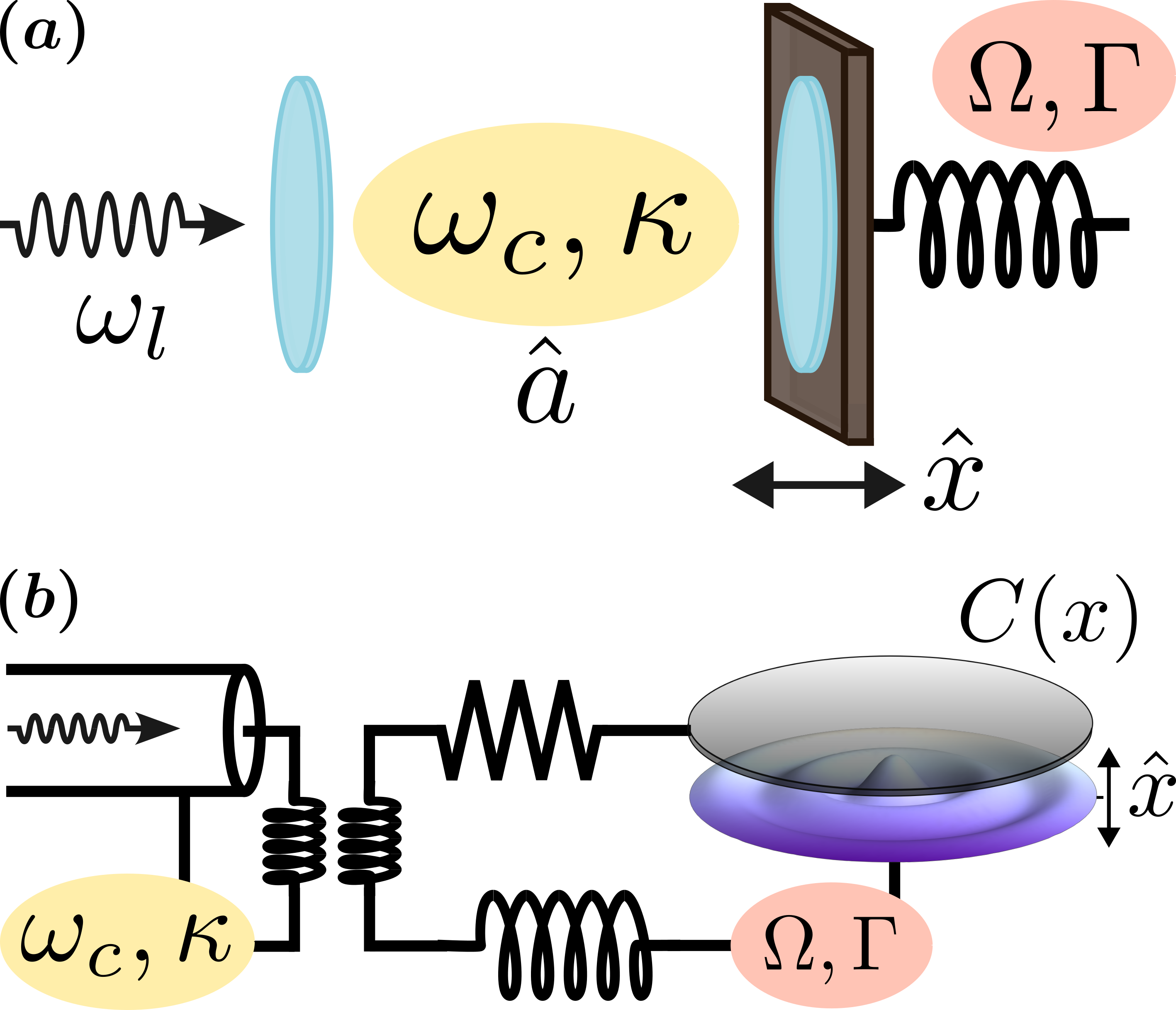}
    \caption{Common examples of cavity optomechanics setups, highlighting the role of the cavity and the mechanical mode. (a) Cavity optomechanics setup consisting of a Fabry--Pérot cavity with a movable end mirror. (b) Cavity optomechanics analogue using electrical components.}
    \label{fig:Optomechanical_platforms}
\end{figure}

Optomechanical oscillators are systems in which a cavity mode is coupled to a mechanical degree of freedom. The canonical setup is that of an optical Fabry--Pérot cavity with a movable end mirror, Fig.\ref{fig:Optomechanical_platforms}-a, yet the general framework is quite broad and also encompasses implementation based on electrical circuits, Fig.\ref{fig:Optomechanical_platforms}-b.

In Hamiltonian formalism, these systems are described as two coupled quantum harmonic oscillators. Using $\hat{a}$ and $\hat{b}$ to denote the annihilation operators for the cavity and the mechanical mode, respectively, their interaction emerges from the Taylor expansion of the cavity resonance frequency $\omega_{c}$ with respect to the mechanical displacement $\hat{x}=x_{\mathrm{zpf}}(\hat{b}^\dagger+\hat{b})$, with $x_{\mathrm{zpf}}=\sqrt{\hbar/2m\Omega}$, the zero point fluctuations. This yields the optomechanical Hamiltonian \cite{Aspelmeyer}:
\begin{equation} \label{eq1_hamiltonian}
    \hat{H} =\hbar\omega_{c}\,\hat{a}^{\dagger}\hat{a}+\hbar\Omega\,\hat{b}^{\dagger}\hat{b}-\hbar G(\hat{a}^{\dagger}\hat{a})\hat{x}.
\end{equation}
The emerging radiation pressure force, $\hat{F}=-\partial \hat{H}/\partial \hat{x}=\hbar G(\hat{a}^{\dagger}\hat{a})$, acts on the mechanical mode resulting in a shift of the optical frequency $\omega_{c}$, directly proportional to the mechanical displacement through $G$, the frequency pull parameter. For the classical cavity field amplitude, $\alpha=\langle \hat{a} \rangle$, and mechanical oscillator displacement, $x=\langle\hat{x}\rangle$, the dynamics--- including dissipation sources--- is described by the following set of coupled equations \cite{Aspelmeyer}:
\begin{equation}\label{eq2_dynamics}
    \begin{split}
        & \dot{\alpha} = \Big[i(\Delta+Gx)-\frac{\kappa}{2}\Big]\alpha+\frac{\kappa}{2}\alpha_{\mathrm{max}}, \\
        & m\ddot{x} = -m\Omega^2x-m\Gamma\dot{x}+\hbar G|\alpha|^2.
    \end{split}    
\end{equation}
Here, $\Delta=\omega_{l}-\omega_{c}$, $\kappa$, and $\alpha_{\mathrm{max}}$ are the detuning of the laser source from the natural resonance frequency, the cavity decay rate, and the maximum light--field amplitude, respectively. The parameters $m, \Omega$ and $\Gamma$ correspond to the mass, the frequency and the intrinsic damping of the mechanical oscillator.

The physics resulting from this coupled system is remarkably rich: when the laser is red--detuned, $\Delta<0$, the mechanical mode is cooled down, whereas blue detuning, $\Delta>0$, leads to negative damping \cite{Kippenberg}.
These processes are captured by the optomechanical induced damping $\Gamma_{\mathrm{Opt}}$, which takes positive or negative values, respectively. When blue--detuned, if:
\begin{equation}
    \Gamma+\Gamma_{\mathrm{Opt}}<0,
\end{equation}
the system enters the nonlinear regime and begins to exhibit self-sustained oscillations, effectively turning into a phase oscillator. Note that, unlike in forced oscillators, the phase here is completely free, \textit{i.e.} it has no reference to an external time. 
If $\Gamma, |\Gamma_{\mathrm{Opt}}| \ll\Omega$, the stable limit cycle is reached after many periods and the dynamics of the mechanical mode can be effectively described introducing amplitude--phase variables $(A, \phi)$:
\begin{equation}\label{eq_amplitude-phase}
        x(t) = A\cos(\phi),
\end{equation}
with $\phi\simeq-\Omega t$. In this case, the light field can be integrated out \cite{MarquardtGirvin} and, introducing $F(t)$ to account for all external forces, the dynamics close to the limit cycle $\bar{A}$ is captured by the \textit{Hopf equations} \cite{collectiveDynamics, Lauter}:
\begin{equation} \label{eq3_OMcell}
    \begin{split}
        & \dot{A} = -\gamma(A-\bar{A})+\frac{F(t)}{m\Omega(A)}\sin{(\phi)}, \\
        & \dot{\phi} = -\bar{\Omega} - \frac{\partial \Omega}{\partial A}\bigg|_{\bar{A}}(A-\bar{A}) + \frac{F(t)}{m\Omega(A) A}\cos{(\phi)}.
    \end{split}
\end{equation}
Here, $\gamma$ is the decay rate that forces the oscillator to remain on the stable orbit, while $\bar{\Omega}$ denotes the steady--state frequency. Notably, this system is typically multistable, admitting several solutions $\bar{A}$ \cite{MarquardtGirvin}, with growing amplitudes as the input laser power increases.

The transition to amplitude--phase variables naturally aligns with the description of phase oscillators and is particularly suitable to analyze collective dynamics and investigate synchronization. 
Furthermore, recasting the optomechanical dynamics as in Eqs.\eqref{eq3_OMcell} is fundamental in this framework, since phases will be treated as the physical node variables.


\begin{figure}[t]
  \centering
  \includegraphics[width=0.48\textwidth]{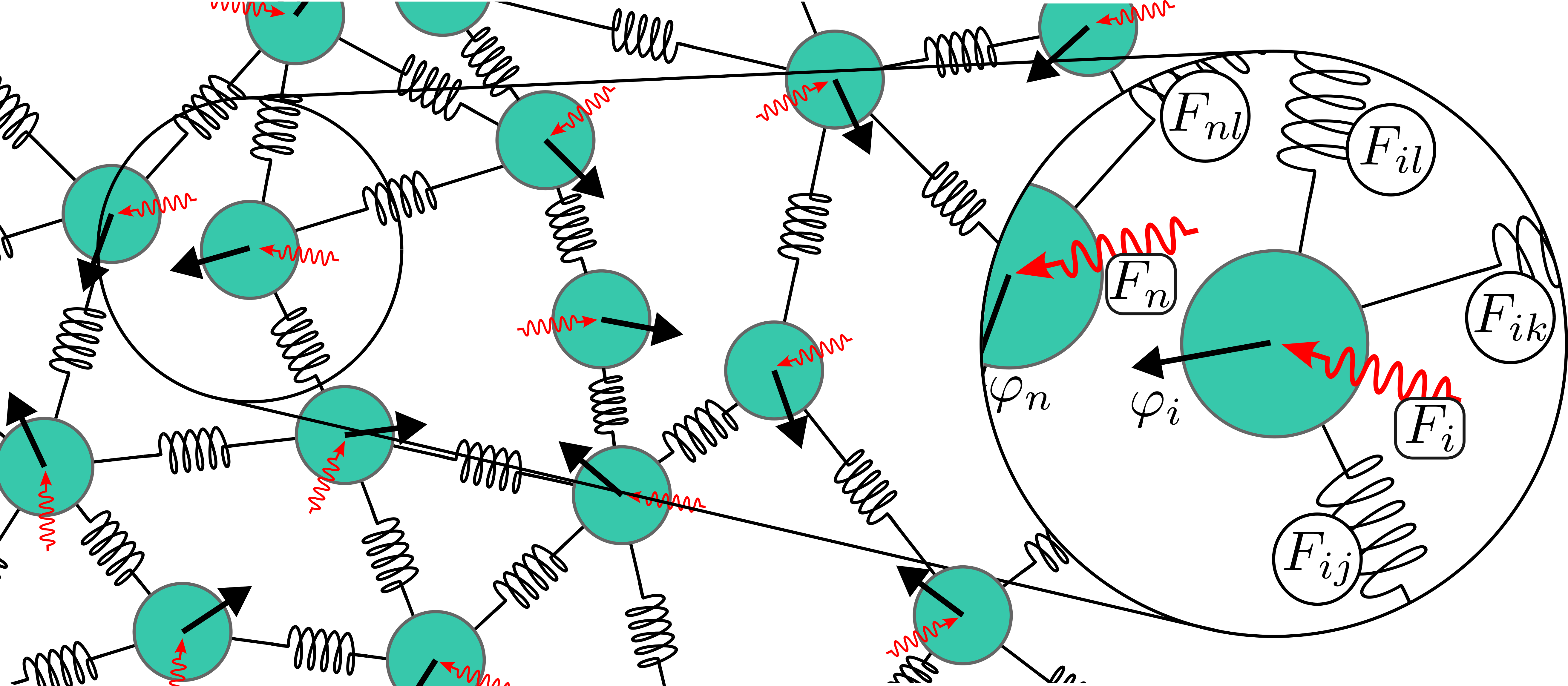}
  \caption{ General view of a physical network assembly with spring--like interactions $F_{ij}$, to reproduce weighted links, and external driving forces $F_i$, for biases.} 
  \label{fig:Network_assembly}
\end{figure}

\subsection{\label{sec_Optomechanical network} Optomechanical network}

The network is composed of optomechanical nodes, labeled by subscripts $i$, arranged to emulate the structure of ANNs.
As illustrated in Fig.\ref{fig:Network_assembly}, this is achieved by introducing spring-like interactions $F_{ij}$ and sinusoidal drive $F_i$ to reproduce the weighted interconnections between nodes and the individual neuron biases. Respectively, these are defined as:
\begin{align}
    &F_{ij} = -k_{ij}x_j, \label{eq_elastic_coupling}\\
    &F_i=f_i\sin\left[\psi_i(t)\right].
    \label{eq_drive}
\end{align}
with $k_{ij}$ being the coupling constant between nodes $i$ and $j$, $f_i$ and $\psi_i(t)$ the intensity and the phase of the drive on node $i$, with the latter defined as $\psi_i(t)=-\omega_{d,i}t+\psi_{o,i}$.

In the regime of weak couplings, namely when force terms are such that $k_{ij}/(m_i\bar{\Omega}_i^2), f_i/(m_i\bar{A}_i\bar{\Omega}_i^2)\ll1$, and, assuming $\gamma_i/\bar{\Omega}_i\ll1, (\bar{A}_i/\bar{\Omega}_i)\partial\Omega_i/\partial A_i \ll1$ with $\omega_{d,i}\simeq \bar{\Omega}_i$, the collective phase dynamics arises integrating out the amplitude fluctuations first.
This yields a set of coupled equations for the node phases,
$\dot{\phi}_i=K_i(\phi_1,\cdots,\phi_N)$, 
where the coefficients depend on the coupling strengths $k_{ij}$ and $f_i$. 
Compared to the already existing effective \textit{Hopf--Kuramoto model} described in \cite{Lauter, Marquardt-Ludvig}, here we derive a richer model due to the inclusion of drives from Eq.\eqref{eq_drive}. Specifically, focusing on the slow dynamics (\textit{i.e.} neglecting all terms with frequencies of order $\Omega$) the phase of each node is given by:
\begin{widetext}
    \begin{equation} \label{eq_opto_net}
    \begin{split}
        \dot{\phi_i} & = -\bar{\Omega}_i+\sum_{j\neq i}\bigg(\frac{\xi_{ij}}{2}\cos{(\phi_j-\phi_i)}\bigg) +\frac{\nu_i}{2}\sin{(\psi_i-\phi_i)} 
        %
        + \frac{\partial\Omega_i}{\partial A_i}\bigg|_{\bar{A}_i}\frac{\bar{A}_i}{2\gamma_i}\bigg[\sum_{j\neq i} \bigg(\xi_{ij}\sin(\phi_j-\phi_i)\bigg)-\nu_i\cos{(\psi_i-\phi_i)}\bigg]  +\\
        & 
        +\sum_{j\neq i}\sum_{k\neq j}\bigg[ \frac{\xi_{ij}\xi_{jk}}{8\gamma_i}\bigg(\sin{(2\phi_j-\phi_k-\phi_i)-\sin{(\phi_k-\phi_i)}}\bigg)\bigg] 
        %
        + \sum_{j\neq i}\sum_{k\neq i} \bigg[ \frac{\xi_{ij}\xi_{ik}}{8\gamma_i}\bigg(\sin{(\phi_j+\phi_k-2\phi_i)}\bigg)\bigg] +\\
        & 
        + \sum_{j\neq i}\bigg[\frac{\xi_{ij}\nu_{j}}{8\gamma_i}\bigg(\cos{(\psi_j-\phi_i)+\cos{(2\phi_j-\psi_j-\phi_i)}}\bigg)\bigg] 
        %
        - \sum_{j\neq i}\bigg[\frac{\xi_{ij}\nu_{i}}{4\gamma_i}\cos{(\psi_i+\phi_j-2\phi_i)}\bigg] -\frac{\nu_{i}^2}{8\gamma_i}\sin{(2\psi_i-2\phi_i)},
    \end{split}
\end{equation}
\end{widetext}
with:
\begin{align}
    \xi_{ij}& =\frac{k_{ij}\bar{A}_j}{m_i\Omega_i\bar{A}_i}, \label{eq_weight_definition}\\
    \nu_{i} & =\frac{f_i}{m_i\bar{\Omega}\bar{A}_i}. \label{eq_bias_definition}
\end{align}

\begin{figure}[b]
  \centering
  \includegraphics[width=0.48\textwidth]{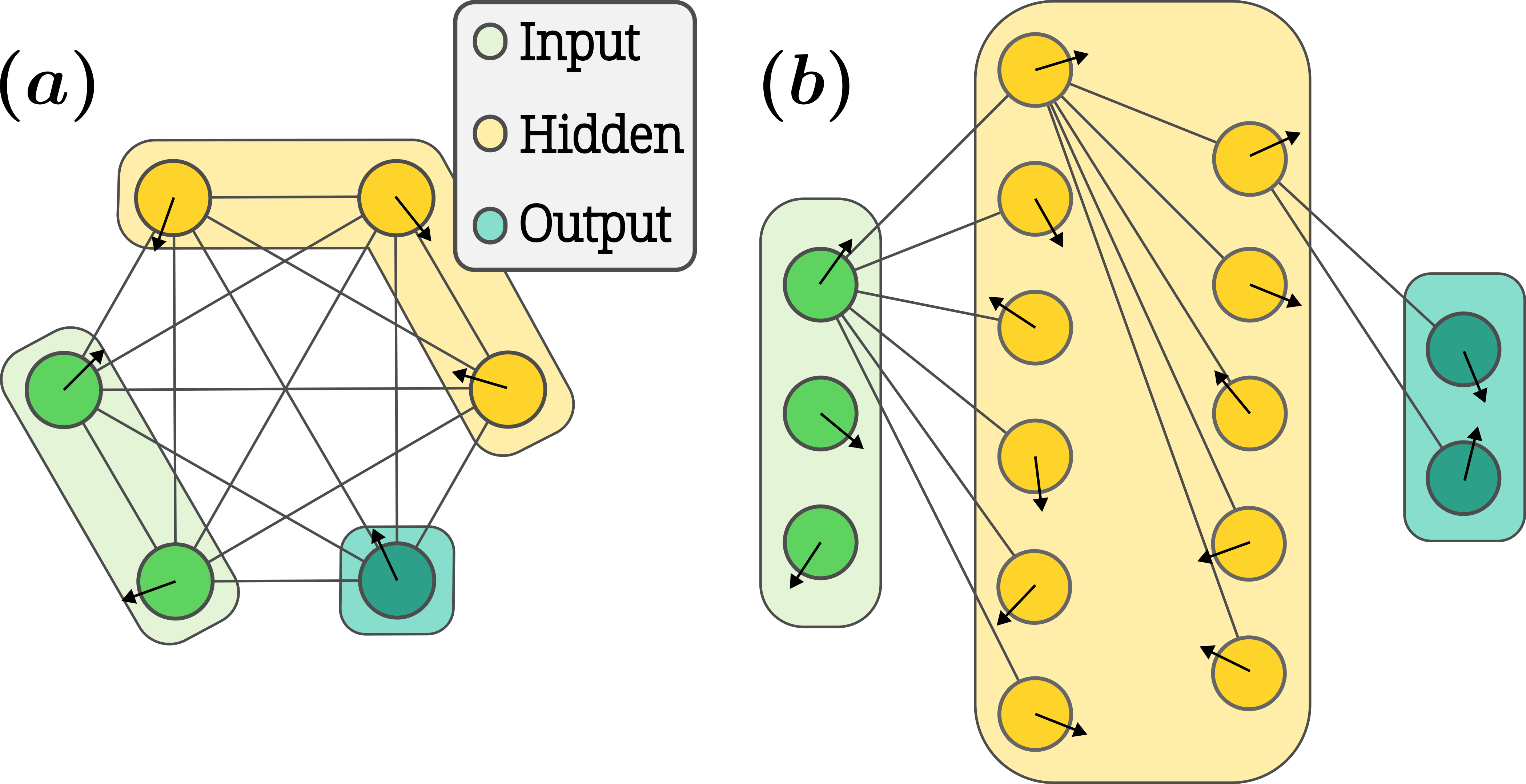}
  \caption{Schematic illustration of the two main architectures for implementing a network in the context of machine learning. (a) Fully connected graph. (b) Multilayer perceptron. Input, hidden and output nodes appear as green, yellow and turquoise colored circles, respectively.}
  \label{fig:network_designs}
\end{figure}

Eq.\eqref{eq_opto_net} is valid in the most general case for fully connected networks (see Fig.\ref{fig:network_designs}.a), however, it can be easily tailored to different designs by properly adapting the span of indices in the sums. This could be the case, for instance, of the multilayer perceptron in Fig.\ref{fig:network_designs}.b.  

Assuming the decay rate is much bigger compared to the other frequencies, namely $\gamma_i\gg\xi_{ij}, \nu_i,\frac{\partial\Omega_i}{\partial A_i}\big|_{\bar{A}_i}\bar{A}_i$, Eq.\eqref{eq_opto_net} can be linearized by retaining only first--order terms, which means that amplitude fluctuations become negligible. 
This is the case, for instance, in drum resonators, as analyzed in Sec. \ref{sec:_Experimental feasibility}.
Note that, to satisfy this requirement, in addition to weak couplings, one shall operate far enough from the Hopf bifurcation as, by definition, $\gamma$ vanishes at the critical point.
%
Ultimately, assuming all natural frequencies are identical, namely $\bar{\Omega}_i=\bar{\Omega}$, one obtains:
\begin{equation}\label{eq_intermediate}
    \dot{\phi_i} = -\bar{\Omega}+\sum_{j\neq i}\bigg[\frac{\xi_{ij}}{2}\bigg(\cos{(\phi_j-\phi_i)} \bigg)\bigg]+\frac{\nu_i}{2}\sin{(\psi_i-\phi_i)},
\end{equation}
which resemble the \textit{Kuramoto model} \cite{Kuramoto}, though with a crucial different: it displays a \textit{cosine} interaction instead of the standard \textit{sine} term. 
As we will see in the next sections, this difference will have a major impact. 
From this point, one might absorb the fast frequency $\bar{\Omega}$ by moving to a rotating frame and introduce the slow phase variable:
\begin{equation}
    \varphi_i=\phi_i+\bar{\Omega}t.
\end{equation}
Therefore Eq.\eqref{eq_intermediate} can be recast into:
\begin{equation}\label{eq_basic_neuro}
    \dot{\varphi_i} = \sum_{j\neq i}\Big[W_{ij}\Big(\cos{(\varphi_j-\varphi_i)}\Big)\Big] +b_i\sin{(\psi_{o,i}-\varphi_i)},
\end{equation}
after renaming the coefficients $\xi_{ij}/2=W_{ij}$ and $\nu_i/2=b_i$.
This final form highlights the analogy with machine learning, where the elements forming the set of trainable parameters $\bm{\theta}$ are typically called weights and biases.
Furthermore, in the present case, $\bm{\theta}$ includes the drive phases $\psi_{o,i}$, \textit{i.e.}:
\begin{equation}\label{eq_trainable_parameters}
    \bm{\theta}=\{W_{ij}(k_{ij}, \bar{A}_{i}),\, b_i(f_i, \bar{A}_{i}),\, \psi_{o,i}\},
\end{equation}
with the dependence on couplings and amplitudes written explicitly to emphasize their relation with adjustable physical quantities. 

Note that, an important distinction for physical implementations arises when examining the node dynamics. In contrast to ANNs--- where each node state is obtained by applying a nonlinear map starting from the inputs--- nodes in optomechanical networks are genuinely dynamical: their states evolve continuously in time and depend on the initial conditions. This feature is completely absent in ANNs, where initial node states are not even defined. One might therefore be tempted to treat the $\varphi_i(t_0)=\varphi_{o,i}$ as additional trainable parameters as well and include them in $\bm{\theta}$. However, as explained in Sec.\ref{Multistability}, doing so can limit the network's operational functionality or hinder the learning process.

%% file: 3_Neural_network_implementation_on_physical_systems.tex
\section{Neural network implementation on physical systems} 

After establishing the fundamental physics underlying an optomechanical network, the focus now shifts on the computational aspect. Conventional ANNs are trained to solve specific tasks through an iterative feedback loop based on a dataset. 
Let $\bm{I}$ denote the set of all inputs ({\em i.e.} training samples). At each iteration, a subset $\bm{\mathfrak{B}}\subseteq\bm{I}$ is selected and fed into the network. If $\bm{\mathfrak{B}}$ coincides with $\bm{I}$, it is referred to as \textit{batch}; otherwise, it constitutes a \textit{minibatch}.
The network then produces a response, namely a set of outputs $\bm{O}(\bm{\mathfrak{B}},\bm{\theta})$ which depends on both the inputs and the trainable parameters $\bm{\theta}$. The quality of these outputs is assessed by comparing them with the target outputs $\bm{O^{\tau}}(\bm{\mathfrak{B}})$ using a scalar function $C(\bm{O}, \bm{O^{\tau}})$, known as \textit{cost} or \textit{loss function}. This function quantifies the deviation between the predicted and the target outputs, with lower values of $C$ indicating higher accuracy.

This first stage, known as \textit{feedforward propagation}, is then followed by the \textit{backpropagation}, which aims to reduce the cost. In this second phase, the partial derivatives of the cost with respect to the trainable parameter, $\partial C/\partial\theta_\alpha$, are computed and used to update the parameters according to the rule:
\begin{equation}\label{eq_update_rule}
    \theta_\alpha\rightarrow\theta_\alpha-\eta\frac{\partial C}{\partial\theta_\alpha},
\end{equation}
where $\eta$ is the \textit{learning rate} and controls the step size of the corrections.
Each cycle, formed by a feedforward and a backpropagation phase, is called an \textit{epoch}. Through successive epochs, the system is expected to improve its performances by reducing the cost, ideally reaching a vanishing value of $C$ once the training is completed. In the following section we describe how this scheme can be implemented on a physical platform, focusing, in particular, on the case of optomechanical networks.

\subsection{\label{sec_Physical_ML} Physical machine learning}
\begin{figure} [t]
    \centering
    \includegraphics[width=0.85\linewidth]{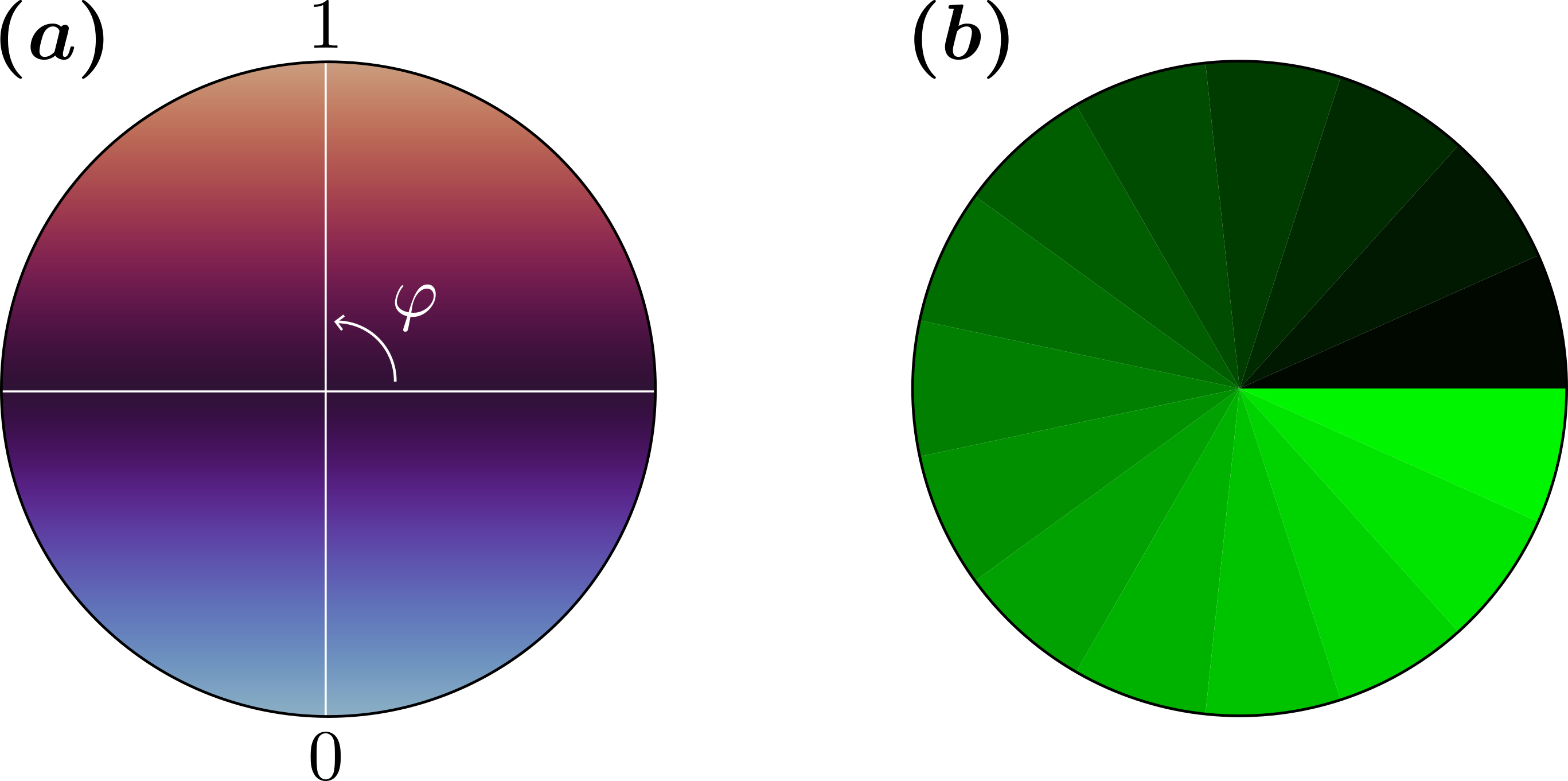}
    \caption{Interpretations of the physical phase angle visualized. (a) Adaptation of a binary variable with the logical states $0$ and $1$ on the opposite side of a circle, separated by $\pi$. (b) Partitioning of the $2\pi$ phase range mapped to different shades of green in the RGB colour system.}
    \label{fig:data_interpretation}
\end{figure}

For any given task, the dataset can be regarded as a map from a set of inputs to a set of target outputs:
\begin{equation}
    \bm{I}\rightarrow \bm{O}^{\tau}. 
\end{equation}
The first step is therefore to adopt a meaningful interpretation of the physical node variables--- here, the phases $\varphi_i$--- to ensure that the system is able to coherently process the dataset. For a binary input--output task, the most natural choice is to associate the logical states with phases separated by $\pi$, using either $\pm \frac{\pi}{2}$ or, similarly, $0$ and $\pi$ (Fig.\ref{fig:data_interpretation}.a). Conversely, when dealing with a finite (discrete or continuous) I--O variables--- such as the intensity of a primary colour in the RGB scales--- the relation can be established by proportionally partitioning the interval $[0, 2\pi[$ in order to reproduce the targeted range of values (Fig.\ref{fig:data_interpretation}.b). 

Secondly, the shape of the network must be designed to suit the task. As for standard neural networks, this implies that the number of physical input and output nodes shall match the dimensions of the vectors within $\bm{I}$ and $\bm{O^{\tau}}$, respectively; while the number of hidden nodes can be chosen freely. Named $\bm{\varphi_{\mathrm{In}}},\bm{\varphi_{\mathrm{Hd}}}$ and $\bm{\varphi_{\mathrm{Out}}}$ the vectors representing the state of input, hidden and output nodes, the full system is described by:
\begin{equation}
    \bm{\varphi}=\{\bm{\varphi_{\mathrm{In}}},\bm{\varphi_{\mathrm{Hd}}},\bm{\varphi_{\mathrm{Out}}}\}.
\end{equation}
However, as the phase of input nodes $\bm{\varphi_{\mathrm{In}}}$ will be fixed by the elements of the batch during the training stage, it is convenient to also introduce the reduced state vector:
\begin{equation}
    \bm{\varphi_{\mathrm{Dyn}}}=\{\bm{\varphi_{\mathrm{Hd}}},\bm{\varphi_{\mathrm{Out}}}\},
\end{equation}
which features only the dynamical nodes.

This process of fixing the inputs mentioned above is known as \textit{data feeding} and, in an optomechanical network, it can be implemented through the use of strong drives. 
According to Eq\eqref{eq_basic_neuro}, input nodes can be controlled considering couplings $b_i\gg W_{ij}$ and, up to small fluctuations, this mechanism locks each $\varphi_i$ to the corresponding $\psi_{o,i}$ (See Appendix~\ref{app_drive_mech} for further details).
Consequently, setting the phases of the input drive vector $\bm{\psi_{o,\mathrm{In}}}=\{\psi_{o,i}\}_{i\in \bm{\mathrm{In}}}$ to match those of an element $\bm{I}_i\in\bm{\mathfrak{B}}$ of the batch, makes it possible to effectively inject the corresponding entry into the physical network.
Thereafter, feedforward propagation spontaneously occurs as the physical network relaxes toward its stationary state $\bm{\varphi^*_{\mathrm{Out}}}$\footnote{The notation used follows the conventions from \cite{Strogatz}}. 
This latter is recorded and the process is repeated for every element of the batch upon forming the set $\bm{O}=\{\bm{\varphi^*_{\mathrm{Out}}}(\bm{I}_i)\}_{i\in\bm{\mathfrak{B}}}$. Finally, the cost is evaluated. Note that, as anticipated, $C$ can generally be any function that quantifies the deviation between the target and the network response; yet, in the present case, it must also respect the $2\pi$--periodicity typical of phase variables.

Next, one must implement the feedback loop. This requires extracting the physical gradient of the cost function, $\partial C/\partial \theta_\alpha$, in order to apply the update rule as in Eq.\eqref{eq_update_rule}. For this purpose, two widely known approaches are Equilibrium Propagation (EP) \cite{Bengio, Laydevant} and Hamiltonian Echo Backpropagation (HEB) \cite{Echo}. 
The system considered here, nonetheless, does not satisfy the necessary conditions to implement either protocol.
In particular, EP is not feasible due to the fact that the dynamics cannot be expressed as a gradient descent of a potential function: \textit{i.e.} it is not possible to find a 
function $U(\bm{\varphi})$ such that $\dot{\varphi}_i = -\partial U/\partial \varphi_i$. 
This is a consequence of the form of the dynamical equations $\dot{\varphi}_i = K_i(\bm{\varphi})$ with: 
\begin{equation}
    \frac{\partial K_i}{\partial\varphi_{j}}\neq\frac{\partial K_j}{\partial\varphi_i},
\end{equation}
due to the presence of  the \textit{cosine} interaction term in 
Eq.\eqref{eq_basic_neuro}.
This indicates that the model possesses a non--reciprocal (rotational) component, which fundamentally alters the system's behavior.
Consequently, the analysis described in \cite{Wang2024} does not apply in this case, and an entirely new training framework is required. 
Note that, the presence of this intrinsic non-reciprocal component also makes it unlikely that one can make progress with any technique like Scattering Backpropagation  \cite{Cin2025}, which in principle sits between EP and HEB, but still typically requires the reciprocity of the dynamics on the level of the linear equations.

Since none of the aforementioned routines could be applied, we developed an alternative training protocol. To this end, a computational scheme was employed: we integrated the dynamics, estimated the cost function, and iteratively computed the partial derivatives $\partial C/\partial \theta_\alpha$, along with the corresponding corrections to the trainable parameters.
This represented a daunting challenge as extracting the physical gradient of a scalar function with respect to the parameters governing the dynamics of the system— here the couplings in the set of differential equations Eq.\eqref{eq_basic_neuro}— is not feasible analytically when an exact solution does not exist. 
Nonetheless, the problem can be tackled numerically by relying on libraries with automatic differentiation. 
In this context, JAX \cite{jax2018github} proved to be particularly effective, providing direct gradient evaluation and supporting highly parallelized simulations via vectorized mappings. This latter, in particular, allowed for the network dynamics to be computed simultaneously for multiple--- or even all--- inputs of the batch.
Further, combined with the Optax module \cite{optax2020github}, JAX also supports the use of optimizers to improve training performances through adaptive learning rates.

When physics-based protocols (like EP and HEB) cannot be employed, relying on a digital approach represents a valuable alternative to gain insights on a physical platform, for example, regarding its expressivity. 

\subsection{\label{Multistability} Multistability}
As training is performed on a real dynamical system, it is important to discuss \textit{multistability}, which has a crucial role in the learning process.

A dynamical system is said to be multistable when it exhibits multiple attractors or stable fixed points, meaning that it can converge to different final states depending on its initial conditions. 
%
Before entering the details, we emphasize that, in our case, multistability might arise because we are using the final stationary state of a dynamical system as the neural network's output.
%
%
This aspect should not be confused with the problem of the nonconvexity of the \textit{cost} function, which represents a distinct issue, already well--known in the field of machine learning.

We begin by noting that multistability can only affect the dynamical nodes of the network. In fact, for input nodes, this behavior is effectively suppressed by the data--feeding mechanism (Sec.\ref{sec_Physical_ML}), which ensures that the state $\bm{\varphi_{\mathrm{In}}}$ remains locked to reproduce a specific entry $\bm{I}_i$, regardless of the initial conditions of the input nodes themselves. Accordingly, in absence of multistability from hidden and output nodes, the network relaxes towards distinct final fixed points depending solely on the injected sample $\bm{I}_i$, namely $\bm{\varphi_{\mathrm{Out}}^*}=\bm{\varphi_{\mathrm{Out}}^*}(\bm{I}_i)$. In this scenario, the training procedure serves two key purposes: it must shape the basins of attraction associated with each input, while ensuring the convergence of the corresponding fixed points to the respective targets, $\bm{\varphi_{\mathrm{Out}}^*}(\bm{I}_i)\rightarrow\bm{\varphi_{\mathrm{Out}}^\tau}(\bm{I}_i)$.
In contrast, when hidden and/or output nodes display multistability, the final state starts depending also on their initial conditions, \textit{i.e.} $\bm{\varphi^*_{\mathrm{Out}}}=\bm{\varphi^*_{\mathrm{Out}}}(\bm{I}_i,\bm{\varphi_{\mathrm{Dyn}}}(t_0))$, resulting in the network producing distinct responses for the same entry. Consequently, the training protocol must also enforce convergence of all solutions associated with a given input $\bm{I}_i$ towards a unique target output, regardless of the initial conditions. Notably, in such cases, the parameter updates $\delta\theta_\alpha$ depend on $\bm{\varphi_{\mathrm{Dyn}}}(t_0)$ as well, and, if the initial conditions vary across epochs, this may lead to mutually incoherent corrections, which must be handled carefully.

Mathematically, the problem can be seen as follows. The total cost $C$ can be decomposed as the average deviation contributed by each element of the batch. Denoting by $C^{(i)}$ the contribution associated with the $i$-th input, the overall cost is given by:
\begin{equation}
    C=\frac{1}{N}\sum_{i=1}^{N}C^{(i)},
\end{equation}
with $N$ denoting the total number of elements in $\bm{\mathfrak{B}}$. Correspondingly, the corrections to the trainable parameters, as given by the update rule in Eq.\eqref{eq_update_rule}, can be similarly expanded in:
\begin{equation}
    \delta\theta_\alpha = -\eta\frac{\partial C}{\partial \theta_{\alpha}}= -\eta\frac{1}{N}\sum_{i=1}^{N}\frac{\partial C^{(i)}}{\partial \theta_{\alpha}}.
\end{equation}
In the spirit of gradient descent, although individual contributions $\partial C^{(i)}/\partial\theta_\alpha$ may point in opposite directions, the high dimensionality of the parameters space provides sufficient flexibility for these contributions to find a compromise and balance out. As a result, despite local conflicts, their average identifies an effective path towards reducing the overall cost function, thereby leading to a minimum.

As anticipated, when the dynamical nodes exhibit multistability, distinct values of $\partial C^{(i,j)}/\partial \theta_\alpha$ may arise for the same input $i$ by varying initial conditions, hence the extra label $j$. 
In such cases, following the same logic as before, one should perform an additional average over many $\bm{\varphi_{\mathrm{Dyn}}}(t_0)^{(j)}$ while keeping the same input, to identify an effective minimization path. This implies adapting the previous update rule to:
\begin{equation}\label{eq_updated_update_rule}
    \delta\theta_\alpha = -\eta\frac{1}{N}\sum_{i=1}^{N}\bigg(\frac{1}{M}\sum_{j=1}^{M}\frac{\partial C_i^{(i,j)}}{\partial \theta_{\alpha}}\bigg),
\end{equation}
where $M$ denotes the total number of initial conditions considered for each input. There is evidence \cite{Wang2024}, obtained within the framework of EP, that this approach allows to train multiple minima simultaneously. 

Adapting the protocol from Eq.\eqref{eq_update_rule} to Eq.\eqref{eq_updated_update_rule} is straightforward, nonetheless, the results we obtained did not meet the expectations as it was not possible to find any parameter configuration to successfully reproduce the XOR. While the causes remain unclear--- though it is likely due to the strong nonlinearity of the system--- the problem can be bypassed by always setting the initial conditions of the dynamical nodes equal to a specific values.
This approach reduces the complexity of the problem but also limits, \textit{a priori}, the operational validity of the final result. In this case, in fact, the correct response of the system is guaranteed only when properly initialized, while its functionality must be verified when starting from states other than the one employed for the training.

From the above considerations, it becomes now clear why including initial phases among trainable parameters is discouraged. Doing so would prevent the implementation of Eq.\eqref{eq_updated_update_rule} and restrict the validity of the final result immediately. Moreover, if the final configuration does not exhibit multistability with respect to some or all dynamical nodes, training their corresponding $\varphi_i(t_0)$ would solely slow down the learning process, as any initial value would lead to the same outcome.


\subsection{\label{Realization of physical interconnections.}Realization of physical interconnections}

Prior to discussing the results obtained regarding the XOR gate, a concrete physical implementation of the network using drum resonators is presented. These devices consist of a mechanically compliant thin membrane, whose vibrational modes depend on its size, tensile stress of the material, and mass density. The membrane's motion can be controlled by adding electrostatic forces through a capacitively coupled gate, typically called top or bottom gate, depending on the geometry. Together, they form a capacitor, with the membrane acting as a \textit{vibrating plate}. When integrated within a microwave cavity circuit  \cite{Teufel, Collin}, this configuration constitute an optomechanical node (Fig.\ref{fig:Optomechanical_platforms}.b) which, in the proper regime, is expected to oscillate according to Eq.\eqref{eq3_OMcell}. 

\begin{figure} [b]
    \centering
    \includegraphics[width=0.95\linewidth]{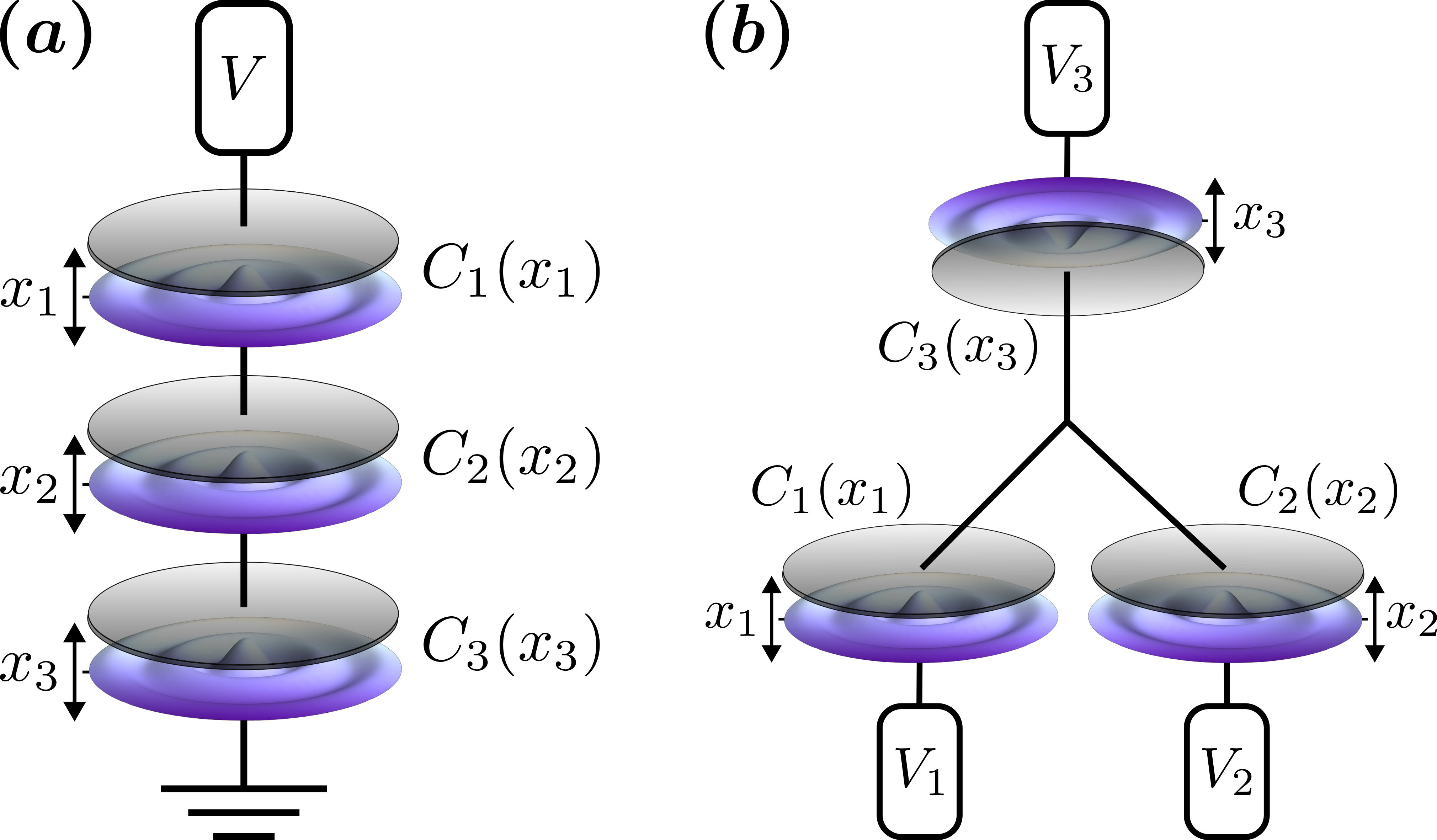}
    \caption{Schemes of the circuits to couple $N=3$ drum resonators. On the left, (a), the \textit{layered topology} (L); on the right, (b), the \textit{all--to--all scheme} (ATA).}
    \label{fig:CC_designs}
\end{figure}

As electical components, they can be integrated within circuits to form networks      \cite{Korppi, Kotler, Lepinay, Youssefi, Chegnizadeh}, enabling the physical implementation of weights $W_{ij}(k_{ij})$ between nodes. Following architectures inspired by ANNs and graph theory, two coupling schemes are considered. These are illustrated in Fig.\ref{fig:CC_designs} in their simplest form, using $N=3$ nodes for clarity, although the designs can be easily extended to an arbitrary number of nodes. The first, shown on the left, corresponds to a \textit{layered topology} (L), while the second, on the right, represents a fully connected or \textit{all--to--all topology} (ATA). The corresponding couplings $k_{ij,\,\mathrm{L/ATA}}$ can be estimated by solving the electrostatic problems presented in Appendix~\ref{app_coupling_derivation}. One finds, respectively:
\begin{align}
    & k_{ij,\, \mathrm{L}} = \Delta V^2(C_{\mathrm{Kir}}^0)^{3}\Big(\frac{1}{C_i^2}\frac{\partial C_i}{\partial x_i}\Big)\Big(\frac{1}{C_j^2}\frac{\partial C_j}{\partial x_j}\Big)\Big|_{x_i,x_j=0}, \label{eq_coupling_L}\\
    & k_{ij,\, \mathrm{ATA}} = -\frac{Q_{i}^0Q_{j}^0}{C_{\Sigma}^0}\Big(\frac{1}{C_i}\frac{\partial C_i}{\partial x_i}\Big)\Big(\frac{1}{C_j}\frac{\partial C_j}{\partial x_j}\Big)\Big|_{x_i,x_j=0}, \label{eq_coupling_ATA}
\end{align}
with the coefficients involved defined as:
\begin{align}
    C_{\mathrm{Kir}} & =\Big(\sum_{i=1}^{N}\frac{1}{C_i}\Big)^{-1},\\
    C_{\Sigma} & =\sum_{i=1}^{N}C_i, \\
    Q_i & = \frac{C_i}{C_{\Sigma}}\Big(\sum_{j=1}^{N}C_j(V_i-V_j)\Big). \label{eq_charges}
\end{align}
Here, variables denoted by the superscript $0$, are evaluated at the rest position, for instance: $C_i^0=C_i(\bm{x}=0)$, with $\bm{x}=\{x_1,\cdots,x_N\}$. 

In both scenarios, the couplings $k_{ij}$ are symmetric and account for all possible pairings. This applies even in the \textit{layered topology} for capacitors that are not directly connected. In the example shown in Fig\eqref{fig:CC_designs}, this is the case for capacitors $C_1$ and $C_3$ which generate a nonzero contribution $k_{13}\neq0$ despite being separated by $C_2$. As a consequence, the resulting optomechanical network is effectively fully connected both in L and ATA. The functional dependence of the $k_{ij}$, however, differs: in each case it is proportional to $1/C_{i}, \partial C_{i}/\partial x_{i}$ and $1/C_{j}, \partial C_{j}/\partial x_{j}$, respectively, but in ATA it also includes the terms $\Delta V_{ij}$ within the $Q_{i}^0$, whereas in L it only features the global voltage drop $\Delta V$. As a result, the ATA design offers greater flexibility, ideally providing $3N-1$ free parameters compared to the $2N+1$ available in L. The theoretical framework is then completed by recalling the definition of weights $W_{ij}$ in Eq.\eqref{eq_weight_definition}, which includes $3$ extra parameters: $m_i, \Omega_i$ an $\bar{A}_i$, for each node. In practice, however, not all the aforementioned quantities are truly tunable. Realistically, only the oscillation amplitudes and the voltage drops can be adjusted, as all the other quantities are basically immutable once drum resonators are built. This reduces the scalability of ATA and L to $2N-1$ and $N+1$, respectively. 
At this point, it is worth noting that the number of links in a fully connected network scales quadratically as $N(N-1)/2$. Consequently, even in networks with relatively few nodes, the number of weights can exceed the number of tunable parameters, thereby limiting the freedom in parameters adjustment for both topologies. However, this constraint does not prevent, \textit{a priori}, to train network of large sizes.

Drawing on these observations, the implementation of the XOR gate was examined on a fully connected ATA topology. In a real platform, the optimal values of the trainable parameters $\bm{\theta}$ can be thus reproduced by carefully adjusting the amplitudes $A_i$ (through the pump intensity) and the voltages $V_i$, together with the drive amplitudes $f_i$ and the phases $\psi_i$.

\subsection{\label{sec:_Experimental feasibility} Experimental parameters range}

In this section we study whether the conditions that have been postulated so far are realizable in a realistic experiment. 
Specifically, we will consider the case of drum resonators. 

The main requirement to derive the reduced model for the phase dynamics is:
\begin{equation}\label{eq_limit}
    \xi_{ij} \ll \gamma,
\end{equation}
which allowed us to discard higher–order terms in $\xi/\gamma$ in Eq.~\eqref{eq_opto_net}.
This condition depends on the coupling between the oscillators and requires careful estimation, since the coupling constant can only be varied within a restricted range of values determined, in the case of drum resonators, by Eqs.~\eqref{eq_coupling_L} and \eqref{eq_coupling_ATA}.
In principle, there is also a constraint on the values of $\nu_i$, but one expects these to be more easily tunable by adjusting the drive intensity.

In order to test the validity of Eq.~\eqref{eq_limit}, we begin by estimating the value of $\gamma$.
This quantity, introduced in Eq.~\eqref{eq3_OMcell}, characterizes the rate at which fluctuations away from the stable orbit $\bar{A}$ decay in time. It results from the competition between mechanical dissipation due to friction and the energy injected by the cavity, captured respectively by $\Gamma$ and $\Gamma_{\mathrm{Opt}}$.
A derivation of the parameter $\gamma$ can be found in Ref.~\cite{Heinrich_PhD}.
For the convenience of the reader, we recall in Appendix~\ref{app_estimation_gamma} the main steps of this derivation, together with additional details regarding the estimation of $\xi_{ij}$.
One finds that $\gamma$ can be written as:
\begin{equation}
    \gamma
    = 
    \frac{\bar{A}}{2}
    \left(
        \frac{\mathrm{d}\Gamma_{\mathrm{Opt}}(A)}{\mathrm{d}A}
    \right)_{\!\bar{A}} .
\end{equation}
Assuming a smooth dependence of $\Gamma_{\mathrm{Opt}}$ on $A$, one expects, on purely dimensional grounds, that:
\begin{equation}
    \gamma = c\big|\Gamma_{\mathrm{Opt}}(\bar{A})\big|,
    \label{eq_gamma=Gamma_dim}
\end{equation}
with $c$ being a factor depending on the input power, hence on the oscillation amplitude $A$.
A more accurate verification of this statement and the estimation of $c$ require the numerical evaluation of $\Gamma_{\mathrm{Opt}}$, which we provide in Appendix~\ref{app_estimation_gamma}.

Since the operating point of the oscillators lies very close to the limit cycle, defined by the condition:
\begin{equation}
    \big|\Gamma_{\mathrm{Opt}}(\bar{A})\big| = \Gamma,
\end{equation}
one therefore expects that:
\begin{equation}
    \gamma = c\,\Gamma.
    \label{eq_gamma=Gamma_final}
\end{equation}
Consequently, to verify the validity of Eq.~\eqref{eq_limit}, one must check whether:
\begin{equation}
    \frac{\xi_{ij}}{\gamma}
    \simeq
    \frac{k_{ij}}{m\Omega}\frac{1}{c\,\Gamma}
    \ll 1,
\end{equation}
holds when inserting the expressions for $k_{ij}$. In the following, we focus on
$k_{ij,\mathrm{ATA}}$, to remain consistent with the rest of the paper, and estimate this ratio using typical experimental parameters.

In the ATA configuration, the coupling constants $k_{ij,\mathrm{ATA}}$ are proportional to the charges $Q_i$ introduced in Eq.~\eqref{eq_charges}, which we approximate as:
\begin{equation}
    Q_i \approx \bar{C}\,\overline{\Delta V},
\end{equation}
where $\bar{C}$ and $\overline{\Delta V}$ denote typical values of the capacitances and voltages, respectively.
This leads to the estimate:
\begin{equation}
    k_{ij,\mathrm{ATA}} 
    \approx 
    \frac{\bar{C}\,\overline{\Delta V}^{\,2}}{N\,\bar{d}^{\,2}},
\end{equation}
where $\bar{d}$ is the typical plate separation.

Combining the above expressions and inserting representative experimental parameters for drum resonators, taken from Ref.~\cite{Xin} and collected in Tab.\ref{tab:SiN drum values}, one finds:
\begin{equation}
    \frac{\xi_{ij}}{\gamma}
    \simeq
    \frac{1}{m\Omega\,c\Gamma}
    \frac{\bar{C}\,\overline{\Delta V}^{\,2}}{N\,\bar{d}^{\,2}}
    \xrightarrow[N=5]{}
    \frac{\overline{\Delta V}^{\,2}}{c}\,(2\times 10^{3}\,\mathrm{V^{-2}}),
\end{equation}
where the voltage $\overline{\Delta V}$ is expressed in volts and $1<c<5$ as found in Appendix~\ref{app_exp_estimation_gamma}.
For voltage biases of the order of a few millivolts—typical in the operating regime of drum resonators—this ratio is much smaller than unity.

Although this estimate should be repeated for any specific experimental configuration, our findings indicate that, within the usual operating range of drum resonators, the condition of Eq.~\eqref{eq_limit} is well satisfied. Consequently, the scheme we propose can be realistically implemented to perform neuromorphic computation in such systems.

\begin{table}[h]
    \caption{\label{tab:SiN drum values} Representative experimental parameters for SiN drum resonators.}
    \centering
    \begin{ruledtabular}
        \begin{tabular}{c c c c}
        & Quantity & Value &  \\
        \hline
        &  $m$ & $10^{-14}$ Kg & \\
        &  $\kappa$ & $10^5$ Hz & \\
        &  $\Omega$ & $6\cdot10^6$ Hz & \\
        &  $\Gamma$ & $6\cdot10^2$ Hz & \\
        &  $G$ & $6\cdot10^{14}$ $\frac{\text{Hz}}{\text{m}}$ & \\
        &  $\Delta=\Omega$ & $6\cdot10^6$ Hz & \\
        &  $\bar{C}$ & $10^{-13}$ F & \\
        &  $\bar{d}$ & $5\cdot10^{-7}$ m & \\
        \end{tabular}
    \end{ruledtabular}
\end{table}




%% file: 4_XOR_Implementation.tex
\section{\label{XOR_implementation}XOR implementation}

\begin{table}[b]
    \caption{\label{tab:XOR_tt} Truth table of the XOR problem.}
    \centering
    \begin{ruledtabular}
        \begin{tabular}{c c c >{\columncolor{gray!15}}c c}
        \multicolumn{5}{c}{XOR gate}\\
        \hline
        & Input1 & Input2 & Output & \\
        \hline
        & 0 & 0 & 0 & \\
        & 0 & 1 & 1 & \\
        & 1 & 0 & 1 & \\
        & 1 & 1 & 0 & \\
        \end{tabular}
    \end{ruledtabular}
\end{table}

Being the simplest nonlinear classification task, the XOR gate provides an ideal validation case. 
This problem has historical relevance in machine learning since, unlike the AND and OR gates, it cannot be represented by a linear threshold function of its inputs and therefore required the introduction of hidden nodes in ANNs. 
In addition, its dataset, listed in Tab.\ref{tab:XOR_tt}, is elementary, containing only four input--output pairs with two--dimensional inputs and one--dimensional outputs. 
We remark here that, to our knowledge, the implementation of a neuromorphic system in terms of optomechanical limit--cycle oscillators considered here represents one of the most complex dynamical systems that have been investigated for this purpose, certainly more complex than coupled spring networks, nonlinear resistance networks, Kerr oscillators, or even Kuramoto model networks. 
Therefore, investigating the training success for even the seemingly simple machine learning task of XOR is highly instructive.

\subsection{\label{Network structure and dataset adaptation}Network structure and dataset adaptation}

\begin{figure} [t]
    \centering
    \includegraphics[width=0.95\linewidth]{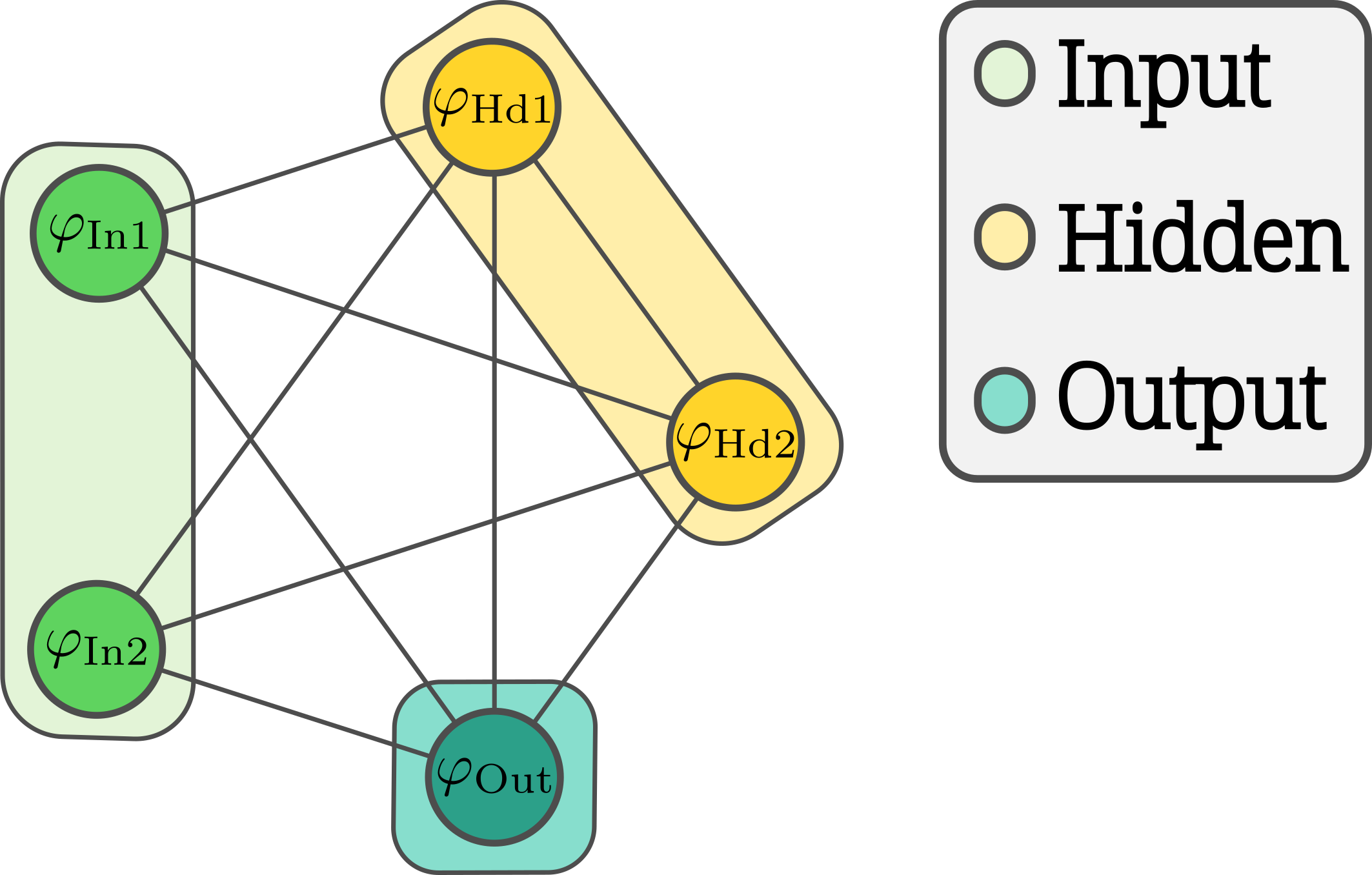}
    \caption{Design used to implement the XOR task. Note the missing connection between the two input nodes.}
    \label{fig:XOR_design}
\end{figure}

Building on the observations made in previous sections, the task is implemented using a fully connected 5--node network, depicted in Fig.\ref{fig:XOR_design}. The number of nodes was chosen to match the minimal size typically required to address this problem using ANNs. The same holds regarding the node partition, which is the following:
\begin{align*}
    \bm{\varphi_{\mathrm{In}}} & = \{{\varphi_{\mathrm{In}1}, \varphi_{\mathrm{In}2}}\},\\
    \bm{\varphi_{\mathrm{Hd}}} & = \{{\varphi_{\mathrm{Hd}1}, \varphi_{\mathrm{Hd}2}}\},\\
    \bm{\varphi_{\mathrm{Out}}} & = \{{\varphi_{\mathrm{Out}}}\},
\end{align*}
with $\bm{\varphi_{\mathrm{Dyn}}}=\{\bm{\varphi_{\mathrm{Hd}}},\bm{\varphi_{\mathrm{Out}}}\}$. The resulting network comprises a total of $15$ trainable parameters $\theta_{\alpha}$, distributed across $9$ weights, $3$ biases, and $3$ initial drive phases $(\psi_{o,i})$:
\begin{align*}
    \bm{W}    = & \{W_{\mathrm{Hd1, In1}},W_{\mathrm{Hd1,In2}},W_{\mathrm{Hd1,Hd2}},W_{\mathrm{Hd1,Out}}, \\
    & W_{\mathrm{Hd2,In1}}, W_{\mathrm{Hd2,In2}}, W_{\mathrm{Hd2,Out}}, W_{\mathrm{In1,Out}}, W_{\mathrm{In2,Out}}\} ,\\
    \bm{b}    = & \{b_{\mathrm{Hd1}},b_{\mathrm{Hd2}},b_{\mathrm{Out}}\} ,\\
    \bm{\psi} = & \{\psi_{\mathrm{Hd1}},\psi_{\mathrm{Hd2}},\psi_{\mathrm{Out}}\}.
\end{align*}
Note that the parameter $W_{\mathrm{In1,In2}}$ is not included among the weights, since the dynamics of input nodes is fully governed by the drives, making its contribution irrelevant. This exclusion allows complete freedom in training weights given that, using the ATA topology, five node amplitudes $\bar{A}_i$ and four voltage differences $\Delta V_{ij}$ can be independently tuned. The binary data are converted in phase variables considering the mapping:
\begin{equation}
    0 \mapsto -\frac{\pi}{2},\;\; 1 \mapsto\frac{\pi}{2}.
\end{equation}

\begin{figure} [t]
    \centering
    \includegraphics[width=0.95\linewidth]{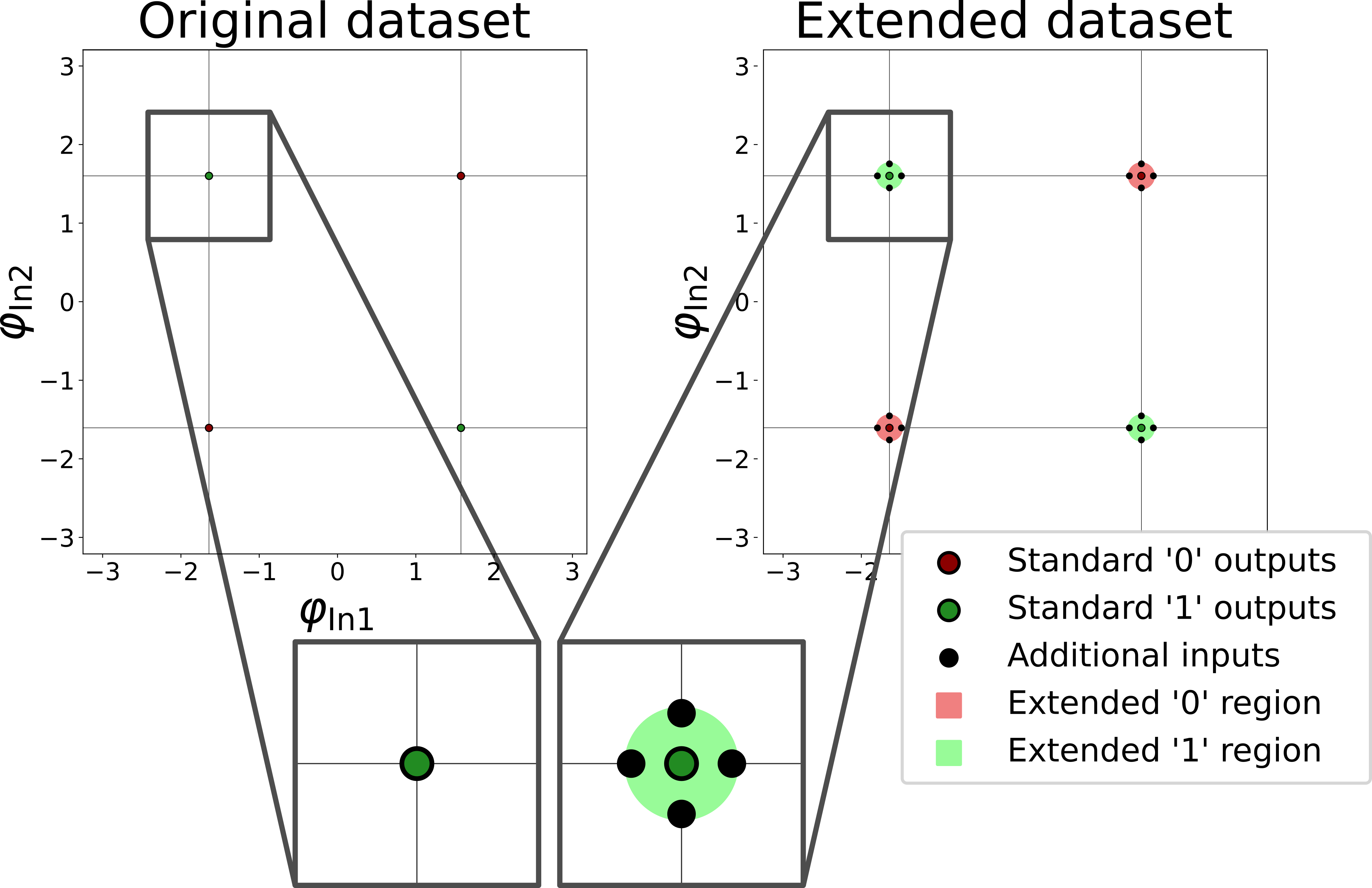}
    \caption{Visualization of the original and extended version of the dataset. Note the shift from having four isolated points to cluster of entries which imply extended regions.}
    \label{fig:dataset_comparison_plot}
\end{figure}

\begin{figure}[t]
    \centering
    \includegraphics[width=0.90\linewidth]{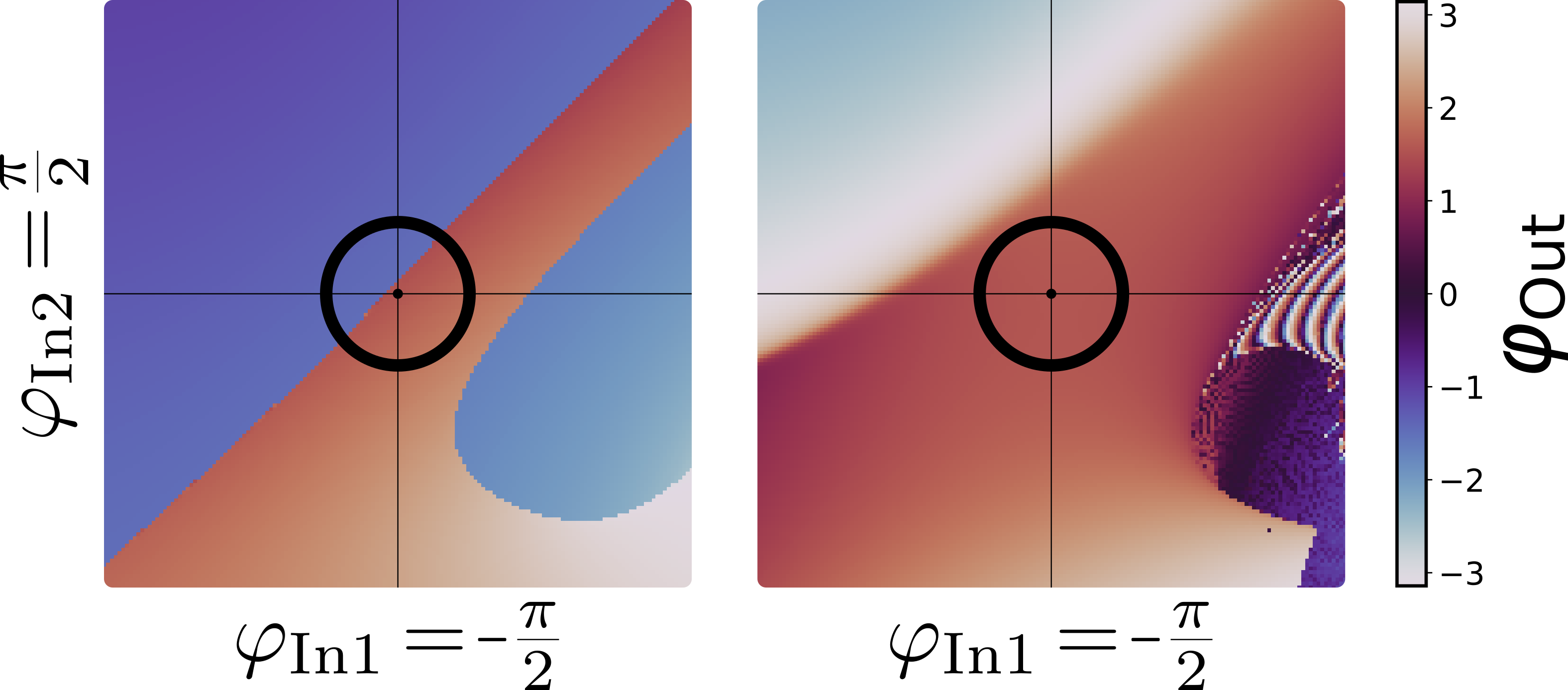}
    \caption{Output response of the system around the $(-\pi/2,\pi/2)$ entry after completing the training. On the left, the final $\varphi_{\mathrm{Out}}$ using only the original truth table's inputs to form the batch. On the right, the result found after training the system on the extended version of the XOR.}
    \label{fig:output_comparison_plot}
\end{figure}

Compared to the datasets of other well--known tasks, such as the Anderson's Iris \cite{iris_53} or the MNIST \cite{lecun2010mnist} problems, the dataset of the XOR consists of isolated points. 
While this characteristic can be easily handled by conventional digital networks, it is less suitable for physical implementations. After training in fact, we observed that correct outputs were often obtained only within narrow regions around the ideal input values, indicating limited tolerance to input variations. Although formally correct, such solutions are not desirable since injecting signals into a real system is inevitably prone to noise, and even small deviations may push the dynamics outside the intended basin of attraction. Therefore, to improve robustness, the dataset was extended from 4 to 20 samples by introducing points in proximity of the original inputs, thereby broadening the training regions. Specifically, shifts of $\delta=\pm0.15$ were applied to each entry, corresponding to an approximate error of $5\%$. Since inputs are 2--dimensional, the comparison can be observed in Fig.\ref{fig:dataset_comparison_plot} where the training areas are plotted against $\varphi_{\mathrm{In}1}$ and $\varphi_{\mathrm{In}2}$ on the $x$-- and $y$--axis. Similarly, improvements in the output response are illustrated in Fig.\ref{fig:output_comparison_plot}, which shows, for $\bm{\varphi_{\mathrm{In}}}=(-\frac{\pi}{2},\frac{\pi}{2})$, the correct response transitioning from being confined to a narrow basin around the entry to a larger region, thus providing a more reliable result.

\subsection{\label{Network dynamics simulations and training loop} Network dynamics and training loop}

Before running simulations and start the training process, the network must be initialized. 
Given that its dynamics is governed by Eq.\eqref{eq_basic_neuro} and that all $W_{ij}$ and $b_i$ are trainable parameters, the system does not possess a fixed time scale, as all frequencies slowly change across epochs. Consequently, the effective time scale is determined by the initial parameters settings. For this reason, these parameters can be randomly initialized by sampling values uniformly distributed within the range $[0,\nu]$
($0\leq W_{ij},b_i\leq\nu$), where $\nu$ is an arbitrary frequency that must satisfy the 
physical constraints $\nu\ll\gamma\ll\Omega$ (regime of validity of the reduced model). Now, since $\nu$ never enters explicitly the equations of motion, rescaling time as $T=t\nu$ makes the weights and biases simply lie within the range $[0,1]$. Ultimately, note that there is no lower bound for $\nu$ which, in principle, can be taken arbitrarily small without altering the physics. As a trade--off, however, the real relaxation time would increase proportionally.
Among the remaining variables, the initial drive phases $\psi_{o,i}$ are uniformly sampled from $[-\pi, \pi]$, whereas the initial dynamical phases $\bm{\varphi_{\mathrm{Dyn}}}(t_0)$ are consistently initialized to a fixed value, that being zero, as motivated in Sec.\ref{Multistability}.


The node dynamics is integrated using a fourth--order Runge--Kutta scheme over a fixed time interval. To enable automatic differentiation through the solver, this latter must be implemented using JAX's \textit{traceable operations}.
Specifically, we wrote a scheme based on JAX-native arrays (which are immutable) and which iterates over time using \textit{jax.lax.scan} instead of a more standard \textit{for}--loop structure. This allows JAX to interpret the dynamics as a computational graph--- a series of elementary operations--- so that gradients can be computed via the chain rule. As a result, both the initial conditions and the system parameters become differentiable variables, enabling optimization with respect to the trainable parameters in $\bm{\theta}$ (for more details on auto--differentiation, the interested readers is invited to read the documentation of JAX at \cite{jax2018github}).

\begin{figure} [t!]
    \centering
    \includegraphics[width=0.87\linewidth]{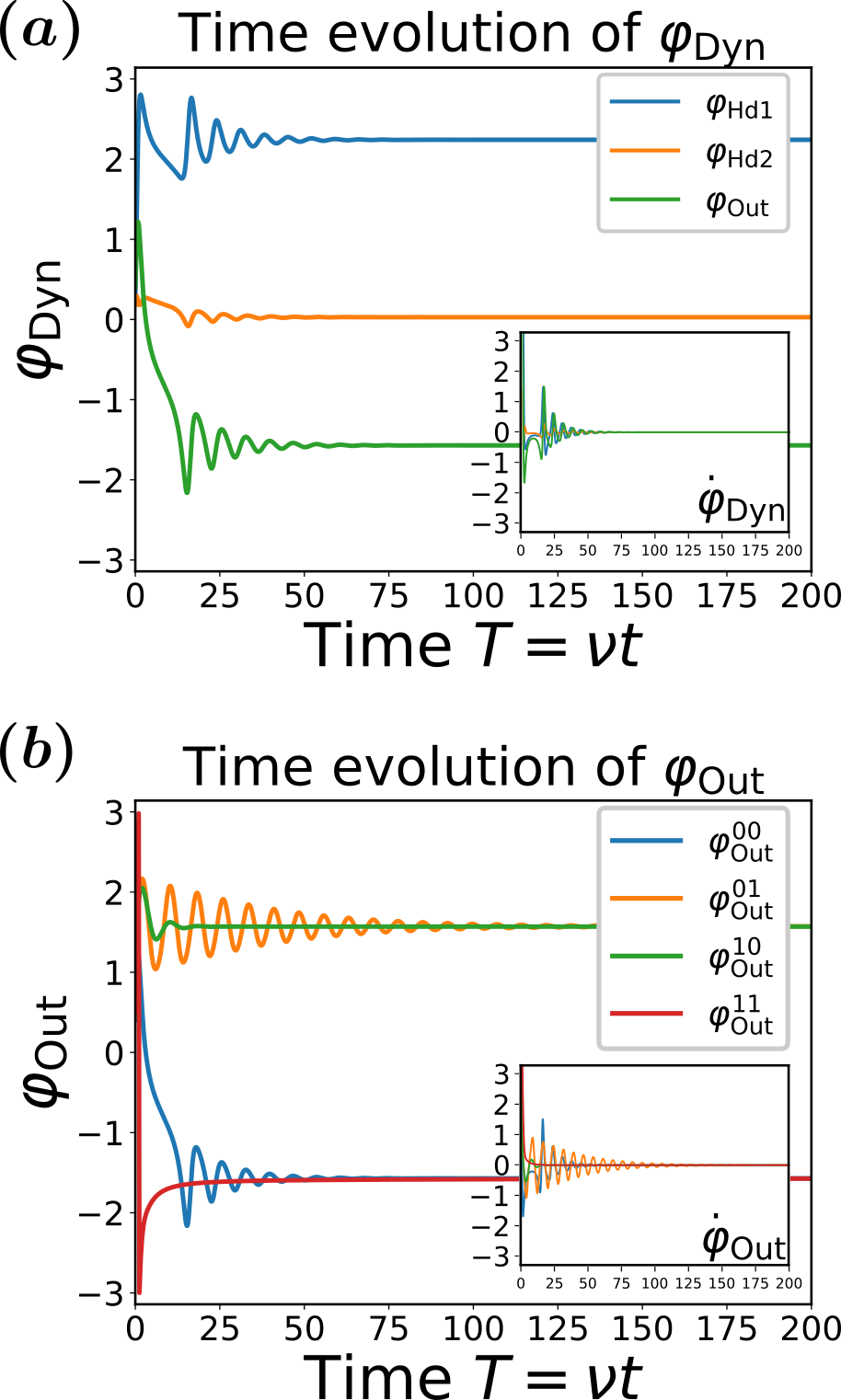}
    \caption{Examples of trajectories reaching stationary states. (a) Time evolution of the dynamical nodes $\bm{\varphi_{\mathrm{Dyn}}}$. Synchronization is reached at around $T\simeq60$. (b) Convergence of $\varphi_{\mathrm{Out}}$ for the four fundamental XOR inputs, labelled with superscripts. The insets show the decay of velocities to zero.}
    \label{fig:time_evolution}
\end{figure}

In most cases, the integration interval considered $(T=200)$ is sufficient to guarantee convergence to a stationary solution, provided one exists. To verify this condition, the angular velocities $\dot{\varphi_i}$ are monitored to check if they vanish during the evolution. Physically, stationarity implies the system has reached a synchronized state, where phase gaps lock to constant values. Notably, such condition increases the reliability of the final output as, otherwise, one would require a protocol to always read the output at the exact same time for every iteration. Two examples of convergence toward synchronization are reported in Fig.\ref{fig:time_evolution}, where the upper panel shows the evolution of the dynamical nodes $\bm{\varphi_{\mathrm{Dyn}}}$, while the lower panel depicts the trajectories traced by the output node $\varphi_{\mathrm{Out}}$ for the four main inputs of the XOR. Insets highlight the decay of angular velocities to zero on the same time interval.

After establishing the numerical scheme, the cost function is defined. This takes the form:
\begin{equation}\label{eq_cost_function}
    C(\bm{O}, \bm{O}^{\tau}) = \sum_{\bm{I}_i\in \bm{\mathfrak{B}}} \frac{1-\cos{\big(\varphi_{\mathrm{Out}}({\bm{I}_i})-\varphi_{\mathrm{Out}}^{\tau}(\bm{I}_i)\big)}}{N_{\mathrm{samples}}},
\end{equation}
where $\varphi_{\mathrm{Out}}({\bm{I}_i})$ is the final output phase obtained injecting the input $\bm{I}_i$, while $\tau$ denotes the corresponding target value. In the present case, the batch $\bm{\mathfrak{B}}$ coincides with the full extended dataset, with the normalization factor therefore fixed at $N_{\mathrm{samples}}=20$. The update of trainable parameters $\bm{\theta}$ was carried out leveraging an optimized version of Eq.\eqref{eq_update_rule}, based on the built--in \textit{grad} function from JAX in combination with the Adam \cite{Adam} module from \textit{Optax}. 
Adam brings together the advantages of momentum averages with those of adaptive learning rates. This smooths the gradient evolution over iterations and facilitates faster and more stable convergence (Explicit formulas can be found in Appendix.~\ref{app_Adam}). 
Note how the use of an optimizer ensures controlled updates to the trainable parameters, allowing their evolution to proceed freely. Ultimately, some weights and biases may exceed $\nu$ at the end of the training, yet this does not pose a problem, as the proper hierarchy of smallness is surely preserved.

When applying this protocol, it was found that the training procedure often struggled to make the cost function vanish completely and became stuck. To address this problem, a cap of $10{,}000$ iterations (epochs) was imposed, after which the routine was restarted from new initial values. This limit was occasionally extended whenever the cost function exhibited an overall decreasing trend. In addition, solutions were retained even if the cost function did not reach zero, provided it decreased below a heuristic threshold set at $C_{\mathrm{crit}}=0.15$. Note that the cost function considered--- Eq.\eqref{eq_cost_function}--- takes values between $[0,2]$.
On average, one such training routine lasts around 10 minutes if run on CPUs.

\subsection{\label{Results}Results}

In this section, the two best results obtained during the training are presented (out of slightly more than $100$ runs with different initial trainable parameter values chosen randomly), hereafter referred to as cases A and B.

\begin{figure} [t]
    \centering
    \includegraphics[width=0.81\linewidth]{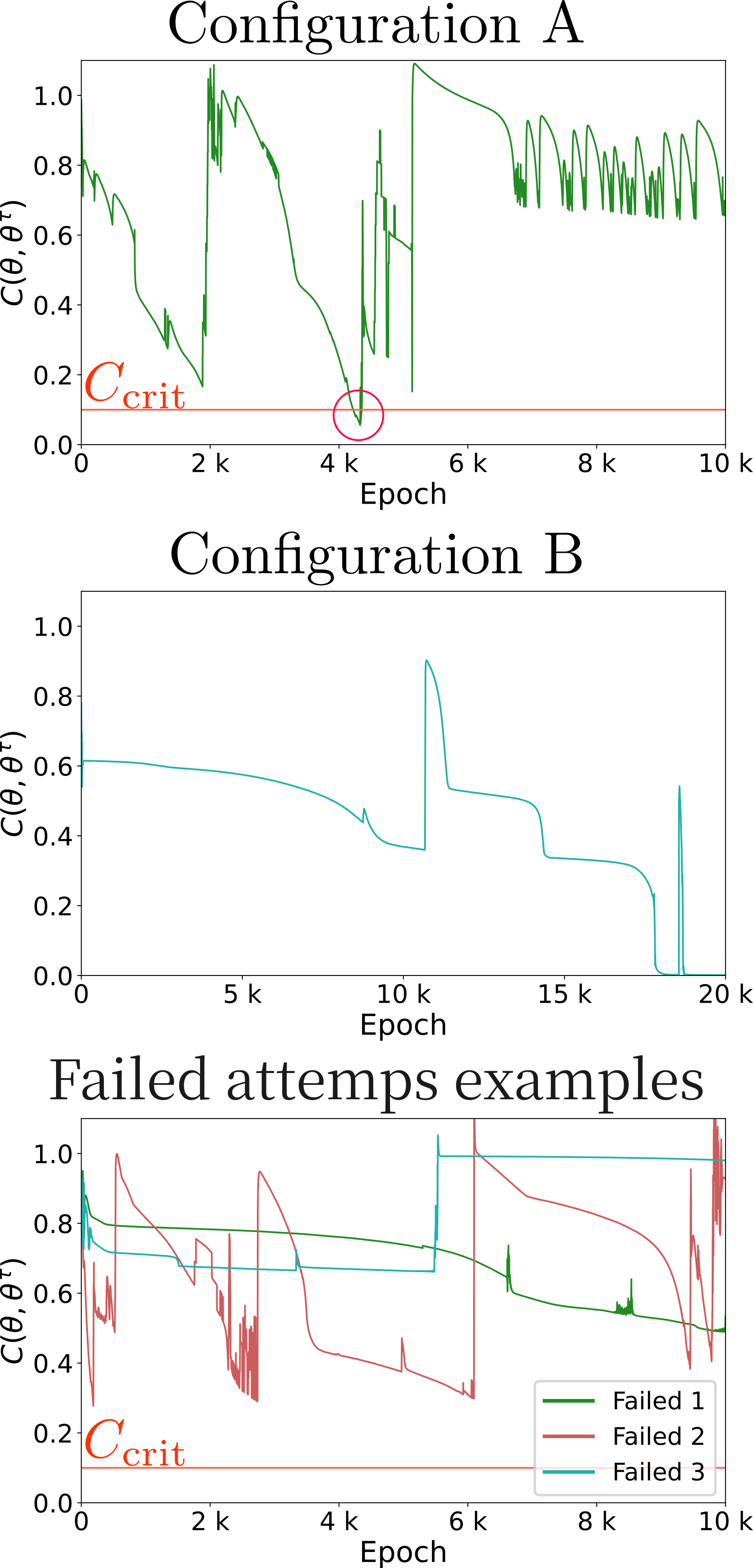}
    \caption{
    Plots of the cost function evolution across epochs for the configurations A and B, with and extra plot displaying some of the failed training attempts. Above) Cost evolution of A: The trend is highly fragmented, characterized by frequent sudden leaps. A value below $C_{\mathrm{crit}}$ was obtained after around $4{,}000$ epochs (circled in red). Middle) Evolution of $C$ for B. Apart from two major jumps, the progression is significantly smoother. A vanishing cost was reached after about $18{,}000$ iterations. Below) 3 examples among the unsuccessful training attempts. As one can observe, the evolution of the cost function is rather unpredictable with $C$ displaying both stiff and irregular trajectories. }
    \label{fig:cost_function_evolution}
\end{figure}

Starting from Fig.\ref{fig:cost_function_evolution}, one can observe the behavior of the cost function across epochs. 
The two configurations exhibit markedly different trends, with B showing a noticeably smoother decay. Yet, in most simulations, the progression more closely resembled that of A, with $C$ evolving in a discontinuous and erratic manner and only rarely dropping below $C_{\mathrm{crit}}$. Within the $100$ runs, this occurred on just four occasions.

The system response can be observed in the density maps shown on the left of Fig.\ref{fig:final_comparison}, where the final output phase $\varphi_{\mathrm{Out}}$ is plotted as a function of the inputs. Both configurations successfully reproduce the XOR within the desired regions when the network is properly initialized at $\bm{\varphi_{\mathrm{Dyn}}}(t_0)=\bm{0}$. The robustness of this result was further tested by monitoring the angular velocities, which confirmed the output node reached a stationary state at least within the black circles. Notably, non--stationary regions can be easily identified on the density maps since they appear as "\textit{blurry areas}". This results from the system still oscillating at $T=200$, and therefore being in a random state when the output is recorded.

\begin{table*}[t]
\caption{\label{tab:optimal_params}Optimal parameters at the end of training for A and B. The values are expressed in units of $\nu$.}
\begin{ruledtabular}
\begin{tabular}{ c c c c c c c c c c }
 & \multicolumn{9}{c}{Weights $\bm{W}$}\\
 & $W_{\mathrm{Hd1, In1}}$ & $W_{\mathrm{Hd1,In2}}$ & $W_{\mathrm{Hd1,Hd2}}$ & $W_{\mathrm{Hd1,Out}}$ & $W_{\mathrm{Hd2,In1}}$ & $W_{\mathrm{Hd2,In2}}$ & $W_{\mathrm{Hd2,Out}}$ & $W_{\mathrm{In1,Out}}$ & $W_{\mathrm{In2,Out}}$\\ 
 \hline
 \rowcolor{gray!15} Case A & $0.68$ & $0.69$ & $0.20$ & $0.63$ & $0.00$ & $0.00$ & $0.83$ & $0.28$ & $0.26$ \\
 \rowcolor{gray!30} Case B & $0.74$ & $0.78$ & $3.05$ & $0.41$ & $0.67$ & $0.64$ & $0.20$ & $0.20$ & $0.20$ \\
 \hline
 & \multicolumn{3}{c}{Biases $\bm{b}$}& & \multicolumn{3}{c}{Psises $\bm{\psi}$}\\
 & $b_{\mathrm{Hd1}}$ & $b_{\mathrm{Hd2}}$ & $b_{\mathrm{Out}}$ & & $\psi_{\mathrm{Hd1}}$ & $\psi_{\mathrm{Hd2}}$ & $\psi_{\mathrm{Out}}$ \\
 \hline
 \rowcolor{gray!15} Case A & $0.76$ & $1.29$ & $0.46$ & \cellcolor{white} & $0.18$ & $3.75$ & $-2.79$ \\
 \rowcolor{gray!30} Case B & $0.00$ & $1.12$ & $0.34$ & \cellcolor{white} & $-1.17$ & $-0.45$ & $-1.14$ \\
\end{tabular}
\end{ruledtabular}
\end{table*}

Finally, the reason to present two configurations becomes clear when analyzing the network's response starting from random initial conditions, that is, when the constraint $\bm{\varphi_{\mathrm{Dyn}}}(t_0)=\bm{0}$ is removed. As anticipated in Sec.\ref{Multistability}, in this case there is no guarantee that the system will consistently reproduce the XOR, since multistability may also affect hidden and output nodes, thereby driving the output phase towards new untrained fixed points. This behavior is clearly observed in the middle of Fig.\ref{fig:final_comparison}, especially for A. Ultimately, both configurations exhibit regions with a speckle pattern but, while for B these are far from the areas analyzed, in A the entries $(1,0)$ and $(0,1)$ becomes completely unreliable. To quantify this discrepancy, the angular mean and variance of the output phase distribution were computed. For each input pair, the dynamics was simulated starting from thirty different initial conditions, and the resulting final states were recorded. Since phases are periodic variables, the ordinary arithmetic average can be misleading; therefore, the angular mean is employed. Each final phase $\varphi_{\mathrm{Out}}^{(i)}$ is associated with a unit vector:
\begin{equation}
    (x_i,y_i) = \Big(\cos{\big(\varphi_{\mathrm{Out}}^{(i)}\big)}, \sin{\big(\varphi_{\mathrm{Out}}^{(i)}\big)}\Big).
\end{equation}
The arithmetic average of these vectors, $(\bar{x},\bar{y})$, then defines the mean phase through:
\begin{equation}
    \bar{\varphi}_{Out} = \arctan{\Big(\frac{\bar{y}}{\bar{x}}}\Big).
\end{equation}
The radius of the mean vector:
\begin{equation}
    R=\sqrt{\bar{x}^2+ \bar{y}^2},
\end{equation}
provides a measure of the dispersion of the phases, with $R=0$ indicating complete dispersion while $R=1$ corresponds to a perfectly peaked distribution at $\bar{\varphi}_{Out}$. The definition for the variance follows as:
\begin{equation}
    V=1-R.
\end{equation}
In the present case, this quantity allows to assess the robustness of the final output when starting from random initial conditions. As it can be observed on the right of Fig.\ref{fig:final_comparison}, within the black circles of B the variance is zero and it only increases in regions that were not considered during the training. On the contrary, in A, areas around the logical pairs $(1,0)$ and $(0,1)$, exhibit a non vanishing variance, compromising the validity of the response. In this case, getting the correct output becomes a matter of chance, corresponding to the system starting from the right basin of attraction.

The optimized parameters of both A and B are reported in Tab.\ref{tab:optimal_params}. 
As a result of the rescaling, these values are expressed in units of the reference frequency $\nu$, introduced in Sec.\ref{Network dynamics simulations and training loop}, and have been rounded to two decimal places for readability. It was verified that this rounding does not affect the final result. 

%% file: 5_Conclusion.tex
\begin{figure*}[t]
    \centering
    \includegraphics[width=0.95\linewidth]{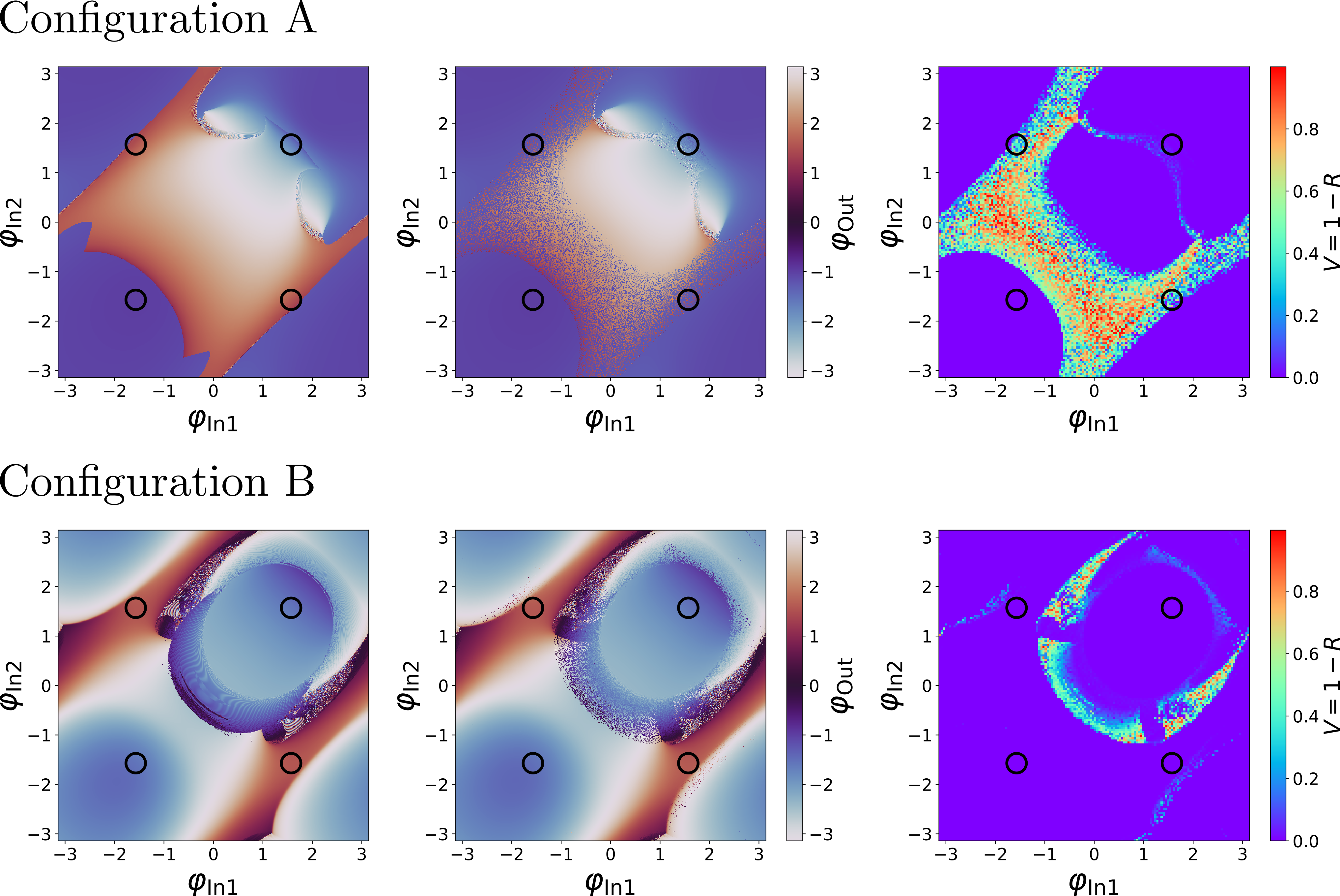}
    \caption{Final response and variance of the two configurations after completing the training. The four circles highlight the areas of the extended XOR dataset. Left)-Middle) Density maps of the final output phase starting from fixed (left) and random (middle) initial conditions. The plots were generated using a grid of $600$ points per side, which can be generated in under a minute using JAX's vector maps. On the left, it is possible to easily distinguish the regions where the system fails to reach a stationary state, which appear as \textit{blurry} areas. Right) Variance maps of A and B. To build the statistics, for each pair of inputs we simulated the dynamics starting from 30 random initial conditions. As a trade--off the maps have a reduced resolution of 100 points per side.
    \label{fig:final_comparison}}
\end{figure*}

\section{Conclusion}

In this work, we have demonstrated that a network of optomechanical oscillators can be employed to perform basic neuromorphic computing tasks, in particular to implement the XOR gate.
Starting from a previously derived microscopic description of the optomechanical oscillator network, we investigated the dynamics of an all-to-all coupled configuration, which could be experimentally realized through controllable electrostatic coupling between oscillators.
We focused on the simplest case in which the system operates deep in the self-oscillation regime, that is, when each optomechanical oscillator is driven by a blue-detuned coherent field away from the Hopf bifurcation.
Under such conditions, the only relevant degree of freedom is the phase of each oscillator which thus enable to encode information and to perform computation via phase synchronization.
However, the system lacks an underlying effective potential in this regime, which prevents the use of EP, a powerful backpropagation protocol, to optimize the trainable parameters. 
We therefore resorted to time-domain simulations of the optomechanical network and found that a system comprising three dynamical nodes (and five nodes overall) already possesses sufficient expressivity to reproduce the XOR task.
The proposed approach is general and can be applied to a wide range of optomechanical networks. Here, we have focused on the case of microwave-driven, drum-like oscillators, for which the inter-node couplings can be tuned either by varying the input power, thereby acting on the amplitude of the stable orbit, or by adjusting gate voltages (see Appendix \ref{app_coupling_derivation}). 
In particular, our estimates suggest that the optimal parameters provided should be reproducible using drum resonators from \cite{Xin}.
While we successfully demonstrated the implementation of the XOR for the network, we also discussed the numerical challenges associated with completing the training procedure. In principle, the method can be extended to more complex tasks; nonetheless, even for a small number of nodes, the protocol become computationally demanding. Therefore, training larger networks surely requires further improvements.

%% file: 6_Acknoledgements.tex
\section*{Acknowledgements}
The authors acknowledge Xin Zhou, from the University of Lille, for fruitful discussions.
We acknowledges financial support from the French Agence Nationale de la Recherche through contract ANR MORETOME ANR--22--CE24--0020--03, and from the French government in the framework of the University of Bordeaux’s France 2030 program/GPR LIGHT.

%% file: 7_Data.tex
\section*{Data availability statement}
The data that support the findings of this article are openly available \cite{Gaspari}, embargo periods may apply.

%% file: appendix.tex
\appendix

\section{
\label{app_drive_mech} Drive on input nodes}
Starting from Eq.\eqref{eq_basic_neuro}:
\begin{equation}
    \dot{\varphi_i} = \sum_{j\neq i}\Big[W_{ij}\Big(\cos{(\varphi_j-\varphi_i)}\Big)\Big] +b_i\sin{(\psi_{o,i}-\varphi_i)},
\end{equation}
we impose $b_i\gg W_{ij}\;\forall j\neq i$ 
for input nodes $\varphi_i\in\bm{\varphi_{\mathrm{In}}}$. Their dynamics can be then recast as:
\begin{equation}\label{eq_input_dynamics}
    \dot{\varphi_i} = b_i\sin{(\psi_{o,i}-\varphi_i)} + \delta_i,
\end{equation}
where $\delta_i$ collects all the small fluctuations due to inter-node couplings. Eq.\eqref{eq_input_dynamics} admits two fixed points:
\begin{gather*}
    \varphi_{i}^*=\psi_{o,i},\\
    \varphi_{i}^*=\psi_{o,i}+\pi,
\end{gather*}
together with their periodic repetitions. Among the two, $\varphi_i^*=\psi_{o,i}$ is the stable solution. Consequently, this is the asymptotic state of the system, since the perturbation term $\delta_i$ prevents the dynamics from locking at the unstable phase solution. 
Expanding around the stable point:
\begin{equation}
    \varphi_i = \varphi_i^*+\epsilon_i=\psi_{o,i}+\epsilon_i,
\end{equation}
one may write:
\begin{equation}
    \begin{split}
        \dot{\epsilon_i} & = b_i\sin{(\psi_{o,i}-\varphi_i)} + \delta_i, \\
        & \approx -b_i\epsilon_i +\delta_i.
    \end{split}
\end{equation}
Solving for $\dot{\epsilon_i}=0$ it enables the estimation of the steady fluctuations:
\begin{equation}
    \epsilon_i \approx \frac{\delta_i}{b_i}.
\end{equation}
In the worst case scenario, a bound for $\delta_i$ is simply given by:
\begin{equation}\
    |\delta_i|=\sum_{j\neq i}|W_{ij}\cos{(\varphi_j-\varphi_i)}|\leq\sum_{j\neq i}|W_{ij}|,
\end{equation}
from which one derives:
\begin{equation} \label{eq_fluctuations}
    |\epsilon_i|\lesssim\frac{\sum_{j\neq i}|W_{ij}|}{b_i}.
\end{equation}
This rough approximation yields a requirement for input locking: $b_i$ must be chosen sufficiently large so that $\sum_{j\neq i}|W_{ij}|/b_i$ is acceptably small. Note that this estimate is highly conservative, as it assumes all phase variables are strongly correlated while decaying toward the same final state. In practice, Eq.\eqref{eq_fluctuations} likely overestimates these fluctuations. From a theoretical perspective, this limitation is irrelevant as we can pick the $W_{ij}$ as small as needed; however, experimentally, relaxing this constraint even slightly provides greater flexibility when modelling the apparatus.

\section{\label{app_coupling_derivation} Framework to estimate the couplings between electronic drum resonators}

To derive the weights $k_{ij,\, \mathrm{L/ATA}}$ for capacitively coupled drum resonators, one must first outline the physics of a capacitor with a movable plate, then threat the case of two such elements connected together, and finally extend the analysis to $N$.

\begin{figure} [h]
    \centering
    \includegraphics[height=4.5cm]{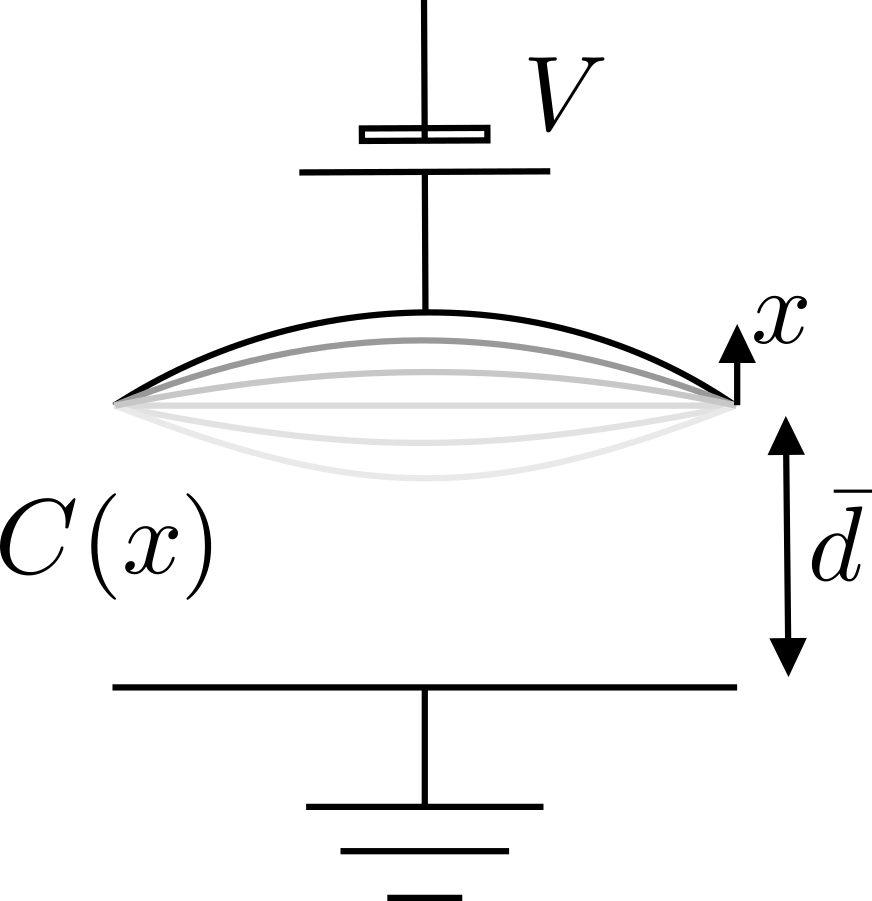}
    \caption{Sketch of a capacitor with a movable plate.}
    \label{fig:movable_capacitor}
\end{figure}
Referring to the sketch in Fig.\ref{fig:movable_capacitor}, a drum resonator can be approximated to a flat oscillating plate. When paired with a second (fixed) plate, it forms a capacitor whose capacitance $C$ depends on the total separation $d$ between the two plates. Denoting by $\bar{d}$ their equilibrium distance (when the resonator is at rest) and by $x$ the displacement of the membrane, one has $C(d)=C(\bar{d}+x)$. The force acting on the movable plate can be obtained computing the variation of electrostatic energy $U_{el}$ stored in the system. In doing so, one must also account for the work done by the voltage source $V$. To this end, the source can be modelled as a capacitor of infinite charge and capacitance, such that $V=Q^*/C^*$ is finite. The total energy then is:
\begin{equation}
    U_{el}=\frac{Q^2}{2C}+\frac{Q^{*2}}{2C^*}.
\end{equation}
Now, consider an infinitesimal, quasi--static displacement $dx$. The corresponding variation in energy is given by:
\begin{equation}\label{eq_el_energy}
    dU_{el}=\frac{Q}{C}dQ-\frac{Q^2}{2C^2}dC+\frac{Q^*}{C^*}dQ^*,
\end{equation}
which must equal the work done by the electrostatic force $F$. Since the total charge $Q+Q^*$ is conserved, one has $dQ=-dQ^*$, and the two contributions cancel, given $Q/C$ is also equal to $V$. Consequently, applying the usual energy--force relation $F=-dU_{el}/dx$, the resulting force is:
\begin{equation} \label{eq_forces}
    F=\frac{Q^2}{2C^2}\frac{d C}{d x}.
\end{equation}
Within the chosen frame of reference in Fig.\ref{fig:movable_capacitor}, negative forces pull in the direction of increasing capacitance. Note that the same outcome can be obtained by differentiating directly the energy stored in $C$:
\begin{equation}
    U_{el}=\frac{Q^2}{2C}=\frac{1}{2}\Delta V^2C,
\end{equation}
while keeping $Q$ constant. Yet this procedure has no physical rationale, because $Q$ is actually changing.

\begin{figure} [ht]
    \centering
    \includegraphics[height=4.5cm]{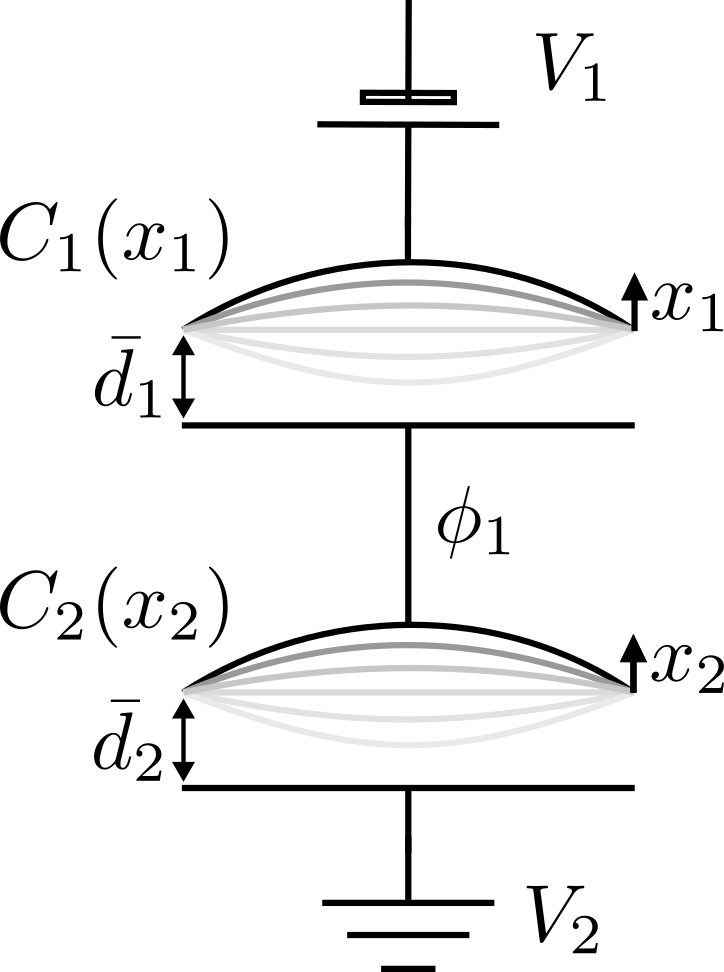}
    \caption{Circuit scheme of two capacitors coupled in series.}
    \label{fig:Double_Capacitor}
\end{figure}
Moving to the case with two capacitors connected in series, as shown in Fig.\ref{fig:Double_Capacitor}, to derive the force, and consequently the coupling, it is necessary to first solve the electrostatic problem. This is outlined by the following set of equations:
\begin{equation}
    \begin{cases}
        (V_1-\phi_1)C_1=Q_1, \\
        (\phi_1-V_2)C_2=Q_2, \\
        Q_1=Q_2=Q,
    \end{cases}
\end{equation}
where, in the present example, $V_2$ is simply connected to ground. By determining the intermediate potential:
\begin{equation}
    \phi_1 = \frac{V_1C_1+V_2C_2}{C_{\Sigma}},
\end{equation}
with $C_\Sigma=C_1+C_2$, the accumulated charges can be expressed as:
\begin{equation}
    \begin{split}
        Q_1 & = \Delta V\frac{C_1C_2}{C_{\Sigma}}=\Delta VC_{\mathrm{Kir}}= Q_2,
    \end{split}    
\end{equation}
where $\Delta V=V_1-V_2$ while $C_{\mathrm{Kir}}=C_1C_2/C_\Sigma$ represents the equivalent capacitance derived from Kirchhoff's law. At this point, proceeding as for the single capacitor case, one finds:
\begin{equation}\label{eq_MC_forces}
    F_i = \frac{Q_i^2}{2C_i^2}\frac{\partial C_i}{\partial x_i}=\frac{1}{2}\Delta V^2C_{\mathrm{Kir}}^2\frac{1}{C_i^2}\frac{\partial C_i}{\partial x_i},
\end{equation}
with $i=1,2$. Since the equivalent capacitance $C_{\mathrm{Kir}}$ depends on both displacements, it follows: $F_i=F_i(x_1,x_2)$. Expanding Eq.\eqref{eq_MC_forces} around the rest positions, one finds: 
\begin{equation}
    \begin{split}
        F_i\simeq\frac{1}{2}\Delta V^2\bigg[&\frac{C_j^2}{C_\Sigma^2}\frac{\partial C_i}{\partial x_i}+ x_i\frac{C_j^2}{C_\Sigma^2}\bigg(\frac{\partial^2C_i}{\partial x_i^2}-\frac{2}{C_\Sigma}\bigg(\frac{\partial C_i}{\partial x_i}\bigg)^2\bigg)+ \\
        +& 2x_j\frac{C_{\mathrm{Kir}}}{C_\Sigma^2}\frac{\partial C_i}{\partial x_i}\frac{\partial C_j}{\partial x_j}\bigg]\bigg|_{x_{i,j}=0},
    \end{split}
\end{equation}
for $i\neq j$. Next, denoting with the superscript $0$ variables computed at rest position, and differentiating with respect to $x_j$, one finds the coupling coefficient as:
\begin{equation}\label{eq_N=2}
    k_{ij}=
    \Delta V^2(C_{\mathrm{Kir}}^0)^3\bigg(\frac{1}{C_i^2}\frac{\partial C_i}{\partial x_i}\bigg)\bigg|_{x_i=0}\bigg(\frac{1}{C_j^2}\frac{\partial C_j}{\partial x_j}\bigg)\bigg|_{x_j=0}.
\end{equation}
One can easily verifies the above result is symmetric. In the following, we tackle the circuit schemes with N capacitors, presented in core of the paper.
\subsection{Layered topology (L)}
\begin{figure} [ht]
    \centering
    \includegraphics[height=6cm]{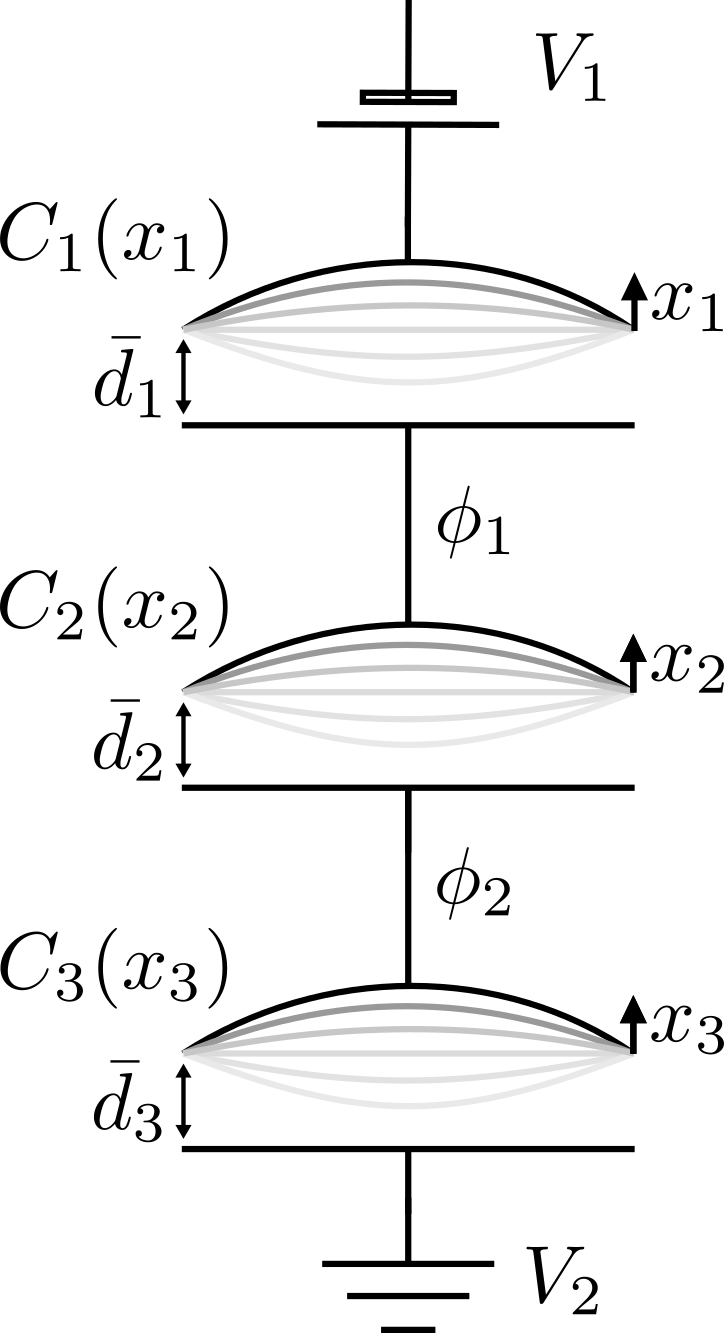}
    \caption{Circuit scheme of three capacitors coupled in series. The generalization to }
    \label{fig:Triple_layered}
\end{figure}
For $N$ drum-like capacitors coupled in series (shown for $N=3$ in Fig.\ref{fig:Triple_layered}) only a few adjustments are needed as the underlying physics remains basically unchanged. In this case, the electrostatic problem is described by:
\begin{equation}\label{eq_L_3_set}
    \begin{cases}
        (V_1-\phi_1)C_1=Q_1, \\
        (\phi_1-\phi_2)C_2=Q_2, \\
        \cdots \\
        (\phi_{N-1}-V_2)C_N=Q_N, \\
        Q_1=Q_2=\cdots=Q_N=Q.
    \end{cases}
\end{equation}
Here, solving for all the intermediate $\varphi_i$ becomes challenging as $N$ grows larger but, by using Kirchhoff's law, it is possible to immediately derive the total charge stored on each capacitor:
\begin{equation}
    Q=\Delta V\Big(\sum_{i=1}^N\frac{1}{C_i}\Big)^{-1}=\Delta VC_{\mathrm{Kir}}.
\end{equation}
Afterwards, replacing this expression in Eq.\eqref{eq_forces} to compute forces, one derives:
\begin{equation}\label{eq_forces_layered}
    F_i = \frac{1}{2}\Delta V^2C_{\mathrm{Kir}}^2\frac{1}{C_i^2}\frac{\partial C_i}{\partial x_i},
\end{equation}
which yields the exact same form of Eq.\eqref{eq_MC_forces}, though indices now span from $i=1$ to $N$. As in the previous case, due to the presence of $C_{\mathrm{Kir}}$, forces depend on the position of every membrane, \textit{i.e.} $F_i=F_i(x_1, x_2, \cdots, x_N)$. This implies that, even though the first and last capacitors are not directly connected, they are nevertheless coupled. In particular, expanding Eq.\eqref{eq_forces_layered} around equilibrium positions, one can make these proportionalities explicit:
\begin{align}
    F_i & =\frac{1}{2}\Delta V^2(C_{\mathrm{Kir}}^0)^2\bigg(\frac{1}{C_i^2}\frac{\partial C_i}{\partial x_i}\bigg)\bigg|_{x_i=0}\bigg[1+ \\
    & \hspace{1cm}+x_i(\cdots)+\sum_{j\neq i}2x_jC_{\mathrm{Kir}}^0\bigg(\frac{1}{C_{j}^2}\frac{\partial C_j}{\partial x_j}\bigg)\bigg|_{x_j=0}\bigg], \notag
\end{align}
Next, differentiating with respect to $ x_j$ for $j\neq i$, and regrouping terms appropriately, one obtains:
\begin{equation}\label{eq_app_coupling_L}
    k_{ij,\,\mathrm{L}} = \Delta V^2(C_{\mathrm{Kir}}^0)^3\bigg(\frac{1}{C_i^2}\frac{\partial C_i}{\partial x_i}\bigg)\bigg|_{x_i=0}\bigg(\frac{1}{C_j^2}\frac{\partial C_j}{\partial x_j}\bigg)\bigg|_{x_j=0}.
\end{equation}
Again, this result matches exactly the one obtained for $N=2$ in Eq.\eqref{eq_N=2} with couplings being symmetrical upon swap of the indices $i\leftrightarrow j$.

\subsection{All--to--all topology (ATA)}
\begin{figure} [ht]
    \centering
    \includegraphics[height=5cm]{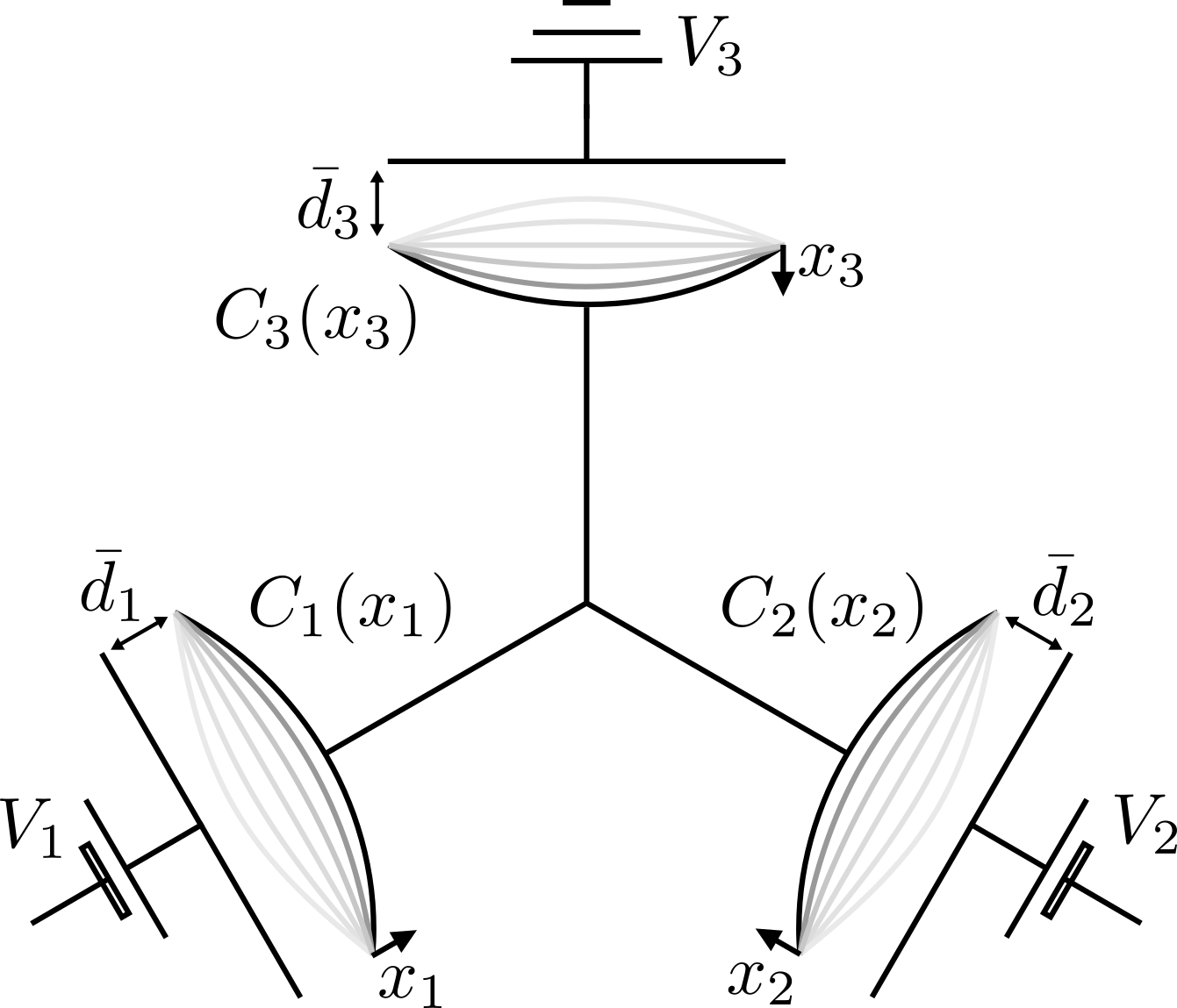}
    \caption{Circuit scheme of three capacitors all--to--all connected.}
    \label{fig:Triple_fullyconnected}
\end{figure}
The fully connected circuit based on $N$ resonators (shown for $N=3$ in Fig.\ref{fig:Triple_fullyconnected}) yields some remarkable differences when computing the couplings. Starting from the set of equations defining the electrostatic problem, one has:
\begin{equation} \label{eq_ATA_3_set}
    \begin{cases}
        (V_1-\phi)C_1=Q_1, \\
        (V_2-\phi)C_2=Q_2, \\
        \cdots \\
        (V_N-\phi)C_N=Q_N, \\
        \sum_{i=1}^N Q_i=0,
    \end{cases}
\end{equation}
and, by summing all equations, one may extract:
\begin{equation}
    \phi=\frac{\sum_{i=1}^{N}C_iV_i}{C_\Sigma},
\end{equation}
with $C_\Sigma=\sum_{i}C_i$. Replacing $\phi$ in Eq.\eqref{eq_ATA_3_set}, enables to find the explicit expressions of the charges stored on each capacitor, namely:
\begin{subequations}
    \begin{align}
        Q_i & = \frac{C_i}{C_\Sigma}\Big[\sum_{j\neq i}C_j(V_i-V_j)\Big],
    \end{align}
\end{subequations}
for $j=1, \cdots, N$. Replacing this result in Eq.\eqref{eq_forces}, results in the force relations:
\begin{equation}
    F_i=\frac{1}{2}Q_i^2\frac{1}{C_i^2}\frac{\partial C_i}{\partial x_i},
\end{equation}
with $F_i$ depending once more on all membrane positions $x_i$, although this time the functional dependence is different compared to Eq.\eqref{eq_forces_layered}. The expansion around equilibrium positions changes coherently:
\begin{equation}
    \begin{split}
        F_i & =\frac{1}{2}\bigg(\frac{Q_i^0}{C_i^0}\bigg)^2\frac{\partial C_i}{\partial x_i}\bigg|_{x_i=0}\bigg[1 +\\
        & + x_i(\cdots) - \sum_{j\neq i} 2x_j\frac{Q_j^0C_i^0}{Q_i^0C_\Sigma^0}\bigg(\frac{1}{C_j}\frac{\partial C_j}{\partial x_j}\bigg)\bigg|_{x_j=0} \bigg],
    \end{split}
\end{equation}
for $j=1, \cdots, N$, and, performing the cross derivatives, one finds:
\begin{equation}\label{eq_app_coupling_ATA}
    k_{ij,\,\mathrm{ATA}}=-\frac{Q^0_iQ^0_j}{C^0_\Sigma}\bigg(\frac{1}{C_i}\frac{\partial C_i}{\partial x_i}\bigg)\bigg|_{x_i=0}\bigg(\frac{1}{C_j}\frac{\partial C_j}{\partial x_j}\bigg)\bigg|_{x_j=0}.
\end{equation}
Again, it is simple to verify that couplings are symmetric.

Comparing the final results, Eq.\eqref{eq_app_coupling_L} and Eq.\eqref{eq_app_coupling_ATA}, it emerges how the all--to--all connectivity grants extra tunable parameters--- the voltages $V_3, \cdots, V_N$--- and is therefore preferable to the layered scheme as it offers greater flexibility. In full generality, the amount of free parameters in L is proportional to $2N+1$, while in ATA it scales as $3N-1$.

\section{\label{app_estimation_gamma}\texorpdfstring{Estimation of $\gamma$}{Estimation of gamma}}

The decay rate $\gamma$ that has been introduced in Eq.\eqref{eq3_OMcell} results from two competing contributions: the intrinsic mechanical damping, $\Gamma$, and the optomechanical energy injection rate, described by $\Gamma_{\mathrm{Opt}}$. 
Their combined action drives the oscillator into a stable limit cycle of amplitude $\bar{A}$, determined by the energy flux balance condition $\Gamma=\Gamma_{\mathrm{Opt}}(\bar{A})$. 
Within this framework, the factor $\gamma$ characterizes the rate at which amplitude perturbations decay, bringing the system back to the stable orbit. 
In this appendix, its derivation is summarized in its key steps, while further details can be found in \cite{collectiveDynamics, Heinrich_PhD}. 

The analysis starts from Eq.\eqref{eq2_dynamics} and the introduction of the amplitude-phase variables $(A, \phi)$, defined as:
\begin{equation}
    \begin{cases}
        x = \bar{x} + A\cos{(\phi)}, \\
        p = A\Omega\sin{(\phi)},
    \end{cases}
\end{equation}
with $\bar{x}$ being the new equilibrium position (displaced by the average radiation pressure force) and $\phi\simeq-\Omega t$. 
These transformations enable to integrate out the cavity field $\alpha(t)$, which takes the form:
\begin{equation}\label{eq_cavity_fields}
    \alpha(t)=e^{i\Psi(t)}\sum_{-\infty}^{\infty}\alpha_ne^{-in\Omega t},
\end{equation}
with:
\begin{align}
    \alpha_{n} & = \alpha_{\mathrm{max}}\kappa\Big(\frac{J_n(\frac{GA}{\Omega})}{\kappa-2i(n\Omega+\Delta+G\bar{x}}\Big),\\
    \Psi(t) & = \frac{GA}{\Omega}\sin{(\Omega t)}.
\end{align} 
As in Eq.\eqref{eq2_dynamics}, here $\alpha_{\mathrm{max}}$ denotes the maximum cavity field amplitude, found solving $\dot{\alpha}=0$ for $\Delta$ and $x$ equal to zero; while the $J_{n}$ are the Bessel's functions of the first kind \cite[ \href{https://dlmf.nist.gov/10.2.E2}{(10.2.2)}]{NIST:DLMF}. Next, one can extract the defining relation for $\bar{x}$ by studying the slow dynamics and, in particular, by checking when the total force balance over one fast oscillation vanishes, \textit{i.e.} solving:
\begin{equation}
    \langle\ddot{x}\rangle_{T}=0
\end{equation}
with $T=2\pi/\Omega$. 
This yields the self-consistent relation:
\begin{equation}
    \begin{split}
        \bar{x} = 
        \frac{\hbar G}{m\Omega^2}\alpha_{\mathrm{max}}^2\kappa^2\sum_{n=-\infty}^{\infty}\frac{J_n(\frac{GA}{\Omega})^2}{\kappa^2+4(n\Omega+\Delta+G\bar{x})^2}.
    \end{split}
\end{equation}
Further, using the amplitude-phase variables, the dynamics can be reshaped into:
\begin{equation}
    \begin{split}
        \dot{A} & = -A\Gamma\sin^2{(\phi)}+\frac{\hbar G}{m\Omega}\sin{(\phi)}|\alpha|^2, \\ 
        \dot{\phi} & =-\Omega -\Gamma\sin{(\phi)}\cos{(\phi)}+\frac{\hbar G}{mA\Omega}\cos{(\phi)}|\alpha|^2.
    \end{split}
\end{equation}
These equations suit better the study of the slow dynamics (averaged over one period $T$) and, in particular, focusing on the amplitude evolution, one finds:
\begin{equation}
    \dot{A} = -\frac{A\Gamma}{2}+\frac{\hbar G}{m\Omega}\langle\sin{(\phi)|\alpha|^2}\rangle_{T},
\end{equation}
which can be recast into:
\begin{equation}\label{eq_amplitude_evolution}
    \dot{A} = -\frac{A\Gamma}{2}-\frac{A\Gamma_{\mathrm{Opt}}(A)}{2},
\end{equation}
introducing:
\begin{equation}\label{eq_gamma_opt}
    \Gamma_{\mathrm{Opt}}(A)=-\frac{2\hbar G}{mA\Omega}\langle\sin{(\phi)|\alpha|^2}\rangle_{T}.
\end{equation}
Note, from Eq.\eqref{eq_amplitude_evolution} one can easily verify the balance condition:
\begin{equation}
    \Gamma=-\Gamma_{\mathrm{Opt}}(\bar{A})
\end{equation}
which holds for all stationary amplitudes $\bar{A}$. Excluding this class of solutions, the value of $\Gamma_{\mathrm{Opt}}$ for any $A$ can be found plugging the definition of the cavity field $\alpha(t)$ into Eq.\eqref{eq_gamma_opt}, which leads to:
\begin{widetext}\label{eq_GammaOpt}
\begin{equation}
    \Gamma_{\mathrm{Opt}}(A) = -\frac{2}{A\Omega}\frac{\hbar G}{m}\alpha_{\mathrm{max}}^2\kappa^2\mathfrak{Im}\sum_{n=-\infty}^{\infty}\frac{J_n(\frac{GA}{\Omega})}{\kappa-2i(n\Omega+\Delta+G\bar{x})}\frac{J_{n+1}(\frac{GA}{\Omega})}{\kappa+2i(n\Omega+\Omega+\Delta+G\bar{x})}.
\end{equation}
\end{widetext}

Ultimately, the defining relation for $\gamma$ arises linearizing Eq.\eqref{eq_amplitude_evolution} in proximity of the limit cycle. Specifically, expanding in terms of $(A-\bar{A})$, the amplitude dynamics becomes:
\begin{equation}
    \dot{A} = -\gamma(A-\bar{A}),
\end{equation}
with $\gamma$ being:
\begin{equation}
    \gamma= \frac{\bar{A}}{2}\Big(\frac{d\Gamma_{\mathrm{Opt}}(A)}{dA}\Big)\Big|_{\bar{A}}.
\end{equation}
Solving the above derivative analytically is a challenging task due to the presence of $\bar{x}$. Nonetheless, retaining this term in symbolic form, enables to write:
\begin{widetext}
    \begin{equation}\label{eq_gamma}
        \gamma =
        \frac{\bar{A}}{2}
        \Big(
        -2\frac{\Gamma_{\mathrm{Opt}}(A)}{A}
        -
        \frac{2}{A\Omega}\frac{\hbar G}{m}\alpha_{\mathrm{max}^2}\kappa^2\mathfrak{Im}\sum_{n=-\infty}^{\infty}
        \Big[        
        \frac{J_n^2-J_{n+1}^2}{D_n(\bar{x})D_{n+1}^*(\bar{x})}
        \frac{G}{\Omega}
        + 2i
        \frac{J_{n}J_{n+1}}{D_n(\bar{x})D_{n+1}^*(\bar{x})}
        \frac{D_{n+1}^*(\bar{x})-D_n(\bar{x})}{D_n(\bar{x})D_{n+1}^*(\bar{x})}\frac{d\bar{x}}{dA}
        \Big]
        \Big)\Big|_{\bar{A}},
    \end{equation}
\end{widetext}

\phantom{here}

\phantom{two}

\noindent where, in the interest of readability, we introduced the following simplifications:
\begin{align}
    & J_n\equiv J_n\Big(\frac{GA}{\Omega}\Big) \hspace{0.5cm}\forall\, n, \\
    & D_n(\bar{x})\equiv\kappa-2i(n\Omega+\Delta+G\bar{x}).
\end{align}

Eq.\eqref{eq_gamma} does not permit the derivation of an universal relation that simply connects $\gamma$ to the other macroscopic parameters. 
As discussed in the main text, by dimensional arguments one expects 
$\gamma \simeq c|\Gamma_{\mathrm{Opt}}(\bar{A})|$.
We consider in the following a specific realization of the model and verify numerically this approximate relation. 
But it is clear that, in order to have a precise estimation, each experimental setup must be treated individually. 


\subsection{\label{app_exp_estimation_gamma}\texorpdfstring{Calculation of $\gamma$ for a specific implementation of drum resonators}{Gamma for SiN drum resonators}}

\begin{figure} [h!]
    \centering
    \includegraphics[width=0.98\linewidth]{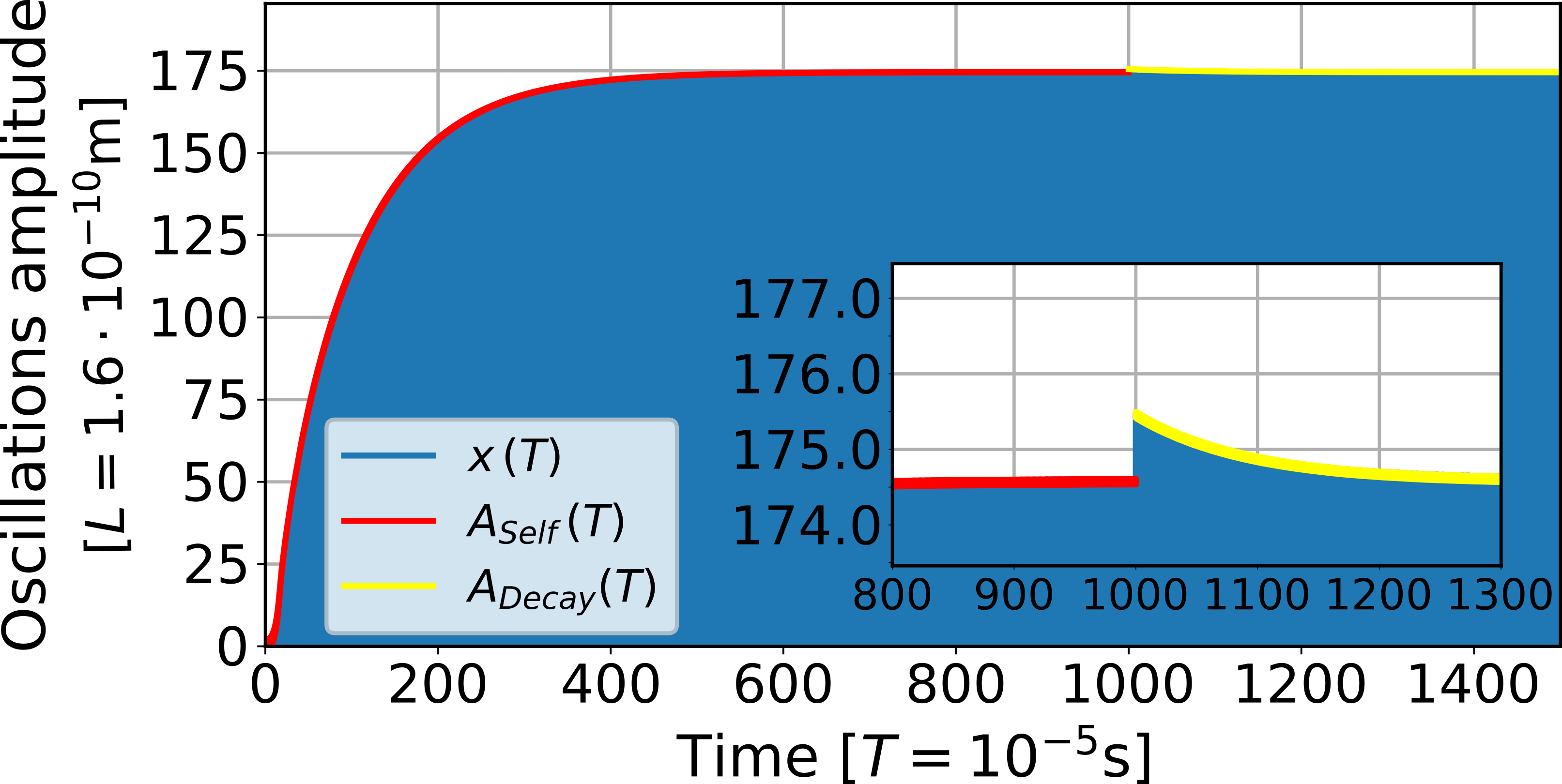}
    \caption{Ring-down time simulation for drum resonators. In blue the actual dynamics of the oscillator (only for positive $x$), whose periods are so short that oscillations cannot be distinguished at this scale. In red the amplitude dynamics which leads to the onset of self-sustained oscillations while, in yellow, the decay of amplitude fluctuations. In the inset, a zoomed view of the "kick" applied to estimate the factor $\gamma$. }
    \label{fig:ring_down}
\end{figure}

In this section we consider the case of 
drum resonators and use the typical parameters of Ref.\cite{Xin} as a concrete case study. 
The estimation of $\gamma$ is achieved either using a purely numerical approach, solving Eqs.\eqref{eq_gamma_opt} and \eqref{eq_gamma} with the sums truncated at an appropriate value of $n$ (in our case, set at $n=10$) or examining ring-down time simulations of the system's decay dynamics.
This latter, in particular, consists in perturbing the stable motion and study how the systems relaxes back. 
An example of this procedure is provided in Fig.\ref{fig:ring_down}, which shows, in the first part, the development of self-sustained oscillations, while, after the perturbation, the decay dynamics.

\begin{figure}[h!]
    \centering
    \includegraphics[width=0.98\linewidth]{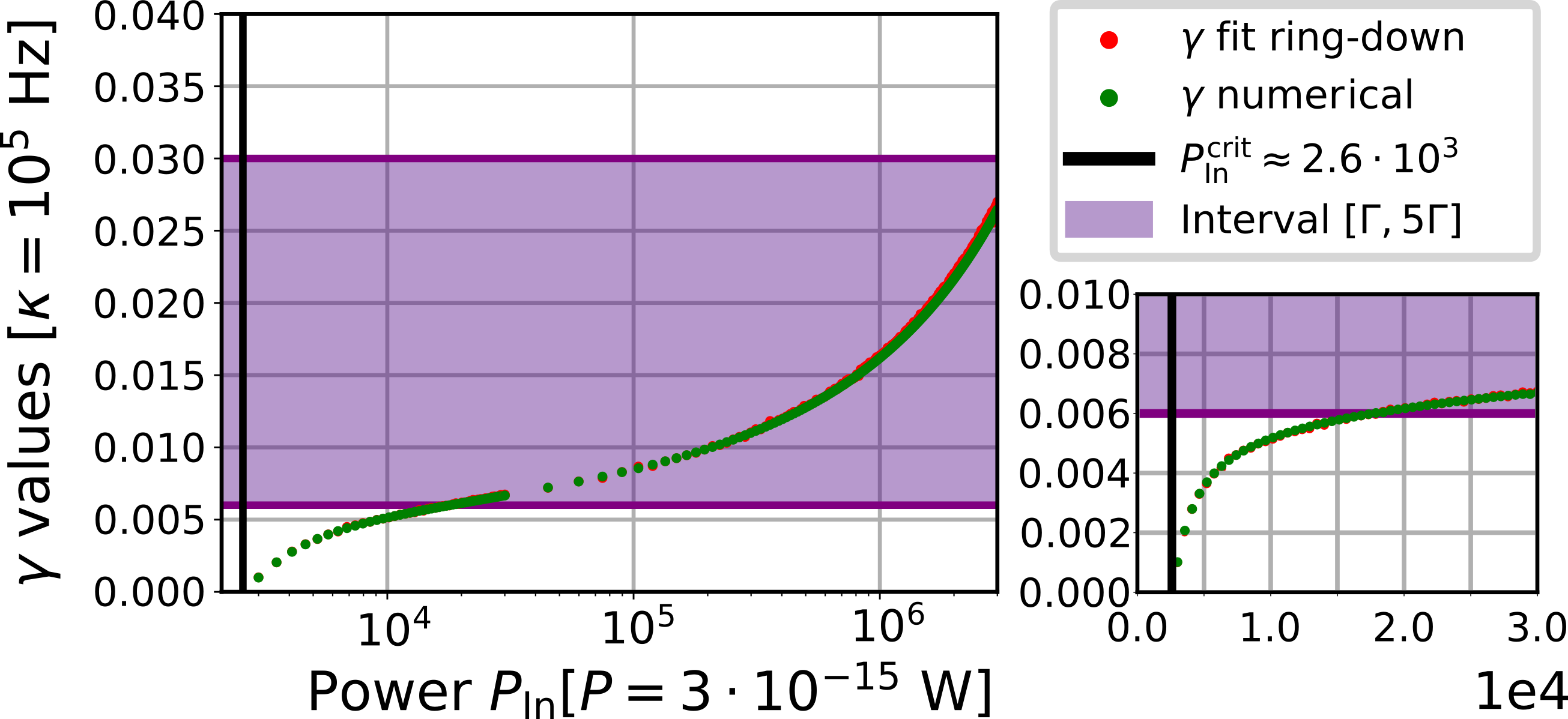}
    \caption{Comparison of the estimations of $\gamma$ for SiN drum resonators \cite{Xin}, obtained using the two approaches presented. In the plots, the purple region highlights the connections with $\Gamma$ which can be formulated as $\gamma\in[\Gamma, 5\Gamma]$. Note, the x-axis on the main graph is in logarithmic scale to provide a clearer picture of the evolution of $\gamma$ across different power scales. On the right, a zoomed view of its behavior when approaching the critical power $P_{\mathrm{in}}^{\mathrm{crit}}$, namely when going closer to the Hopf bifurcation point.}
    \label{fig:fits}
\end{figure}

In the case of drum resonators, the results found using both methods yields highly consistent estimates of $\gamma$. As shown in Fig.\ref{fig:fits}, the agreement remains nearly perfect across the entire power range analyzed, even though realistic pumping powers are on the order of nW. Notably, in both cases, the characteristic behavior of $\gamma$ as the system approaches the Hopf bifurcation is recovered, with $\gamma$ tending to zero at the transition. 

\section{\label{app_Adam}\texorpdfstring{Adam optimizer}{Adam}}
For completeness, the key formulas for implementing the Adam optimizer are listed below.

At its core, Adam traces the exponentially decaying moving averages of the gradient $(\bm{m}_{t})$ and its squared values $(\bm{v}_t)$, defined as:
\begin{align}
    \bm{m}_t & = \beta_1\bm{m}_{t-1}+(1-\beta_1)(\bm{\nabla}_{\hspace{-0.05cm}\theta}\,C), \\
    \bm{v}_t & = \beta_2\bm{v}_{t-1} + (1-\beta_2)(\bm{\nabla}_{\hspace{-0.05cm}\theta}\,C)^2.
\end{align}
Nonetheless, note that, as the starting elements $\bm{m}_0$ and $\bm{v}_0$ are initialized as (vectors of) zeros, the algorithms actually relies on their bias-corrected estimates:
\begin{align}
    \hat{\bm{m}_t} & = \frac{\bm{m}_t}{1-\beta_1^t},\hspace{1cm} \hat{\bm{v}_t} = \frac{\bm{v}_t}{1-\beta_2^t},
\end{align}
and computes the updates according to:
\begin{equation}
    \Delta\bm{\theta}=-\eta\frac{\hat{\bm{m}_t}}{\sqrt{\hat{\bm{v}_t}}+\epsilon}.
\end{equation}
As anticipated in the main text, by leveraging exponentially decaying averages of both the gradient and its square, Adam not only produces adjustable learning rates for each parameter but also introduces numerical stability into the optimization routine. This dual ability ensures better performances while maintaining computational efficiency, as the method relies solely on the first-order gradient $(\bm{\nabla}_{\hspace{-0.05cm}\theta}\,C)$ without the need for second-order derivatives.
In the present work we considered a very standard set of hyperparameters, with $\beta_1=0.9$, $\beta_2=0.999$, $\epsilon=10^{-8}$ and $\eta=10^{-2}$.
\newpage